\documentclass[onecolumn]{ieeetran}

\usepackage{setspace}

\usepackage{cite}
\usepackage{amsmath,amssymb,amsfonts}
\usepackage{algorithmic}
\usepackage{graphicx}
\usepackage{textcomp}

\usepackage{cite}
\usepackage{amsmath,amssymb,amsfonts}
\usepackage{algorithmic}
\usepackage{tabularx} % 在导言区添加

\usepackage{graphicx}
\usepackage{textcomp}
\usepackage{multirow}
\usepackage{hyperref}
\usepackage{epstopdf}
\usepackage{subcaption}
\usepackage{tabularx}
\usepackage{color}
\usepackage{array}
\usepackage{booktabs}
\usepackage{stfloats}
\usepackage{url}
\usepackage{verbatim}
\usepackage{makecell}
\usepackage{multirow}
\usepackage{threeparttable}
\usepackage{subcaption} 
\usepackage{float}
\usepackage{xcolor}
\def\BibTeX{{\rm B\kern-.05em{\sc i\kern-.025em b}\kern-.08em
    T\kern-.1667em\lower.7ex\hbox{E}\kern-.125emX}}
\usepackage{fancyhdr} % 引入页眉页脚管理包

\fancypagestyle{firstpage}{
    \fancyhf{} % 清空所有默认的页眉页脚
    \fancyhead[C]{ % 设置居中页眉
        \parbox{\textwidth}{ \centering \small % 使用 parbox 确保长文本能自动换行，并居中
            T. Zhang, S. Sun, M. Tao, Q. Zhu, and R. Gao, ``Diffuse scattering measurements and mechanism analysis at 8, 12, and 28 GHz for typical building surfaces,'' \textit{npj Wireless Technology}.
        }
    }
}
\begin{document}
\title{Diffuse Scattering Measurements and Mechanism Analysis at 8, 12, and 28 GHz for Typical Building Surfaces}

\author{
    \IEEEauthorblockN{ Tongjia Zhang$^{1}$, Shu Sun$^1$$^{\dag}$, Meixia Tao$^1$, Qiuming Zhu$^2$, Ruifeng Gao$^{3, 4}$}

    \IEEEauthorblockA{$^1$School of Information Science and Electronic Engineering, Shanghai Jiao Tong University,}
    
    \IEEEauthorblockA{Shanghai 200240, China}
    
    \IEEEauthorblockA{$^2$College of Electronic and Information Engineering, Nanjing University of Aeronautics and Astronautics, }
   \IEEEauthorblockA{Nanjing 211106, China}
   
    \IEEEauthorblockA{$^3$ School of Transportation and Civil Engineering, Nantong University, Nantong 226019, China}
    
    \IEEEauthorblockA{$^4$ Nantong Research Institute for Advanced Communication Technologies, Nantong 226019, China}
    
    \IEEEauthorblockA{$^\dag$Corresponding author: Shu Sun (email: shusun@sjtu.edu.cn)}
 }

% \author{
% % Tongjia Zhang, \IEEEmembership{Student Member, IEEE}, 
% % Shu Sun, \IEEEmembership{Senior Member, IEEE},
% % Meixia Tao, \IEEEmembership{Fellow, IEEE},
% % Qiuming Zhu, \IEEEmembership{Senior Member, IEEE}
% % Ruifeng Gao, \IEEEmembership{Member, IEEE}
% Tongjia Zhang, 
% Shu Sun, 
% Meixia Tao, 
% Qiuming Zhu, 
% Ruifeng Gao

%  % \thanks{This work is supported in part by the National Natural Science Foundation of China under Grants 62271310 and 62431014 and in part by the Fundamental Research Funds for the Central Universities of China.}

% \thanks{T. J. Zhang, S. Sun, M. X. Tao are with the School of Information Science and Electronic Engineering, Shanghai Jiao Tong University, Shanghai 200240, China
%  (e-mail: {Mir2712,
% shusun, mxtao}@sjtu.edu.cn).}
% \thanks{Q. M. Zhu is with the College of Electronic and
% Information Engineering, Nanjing University of Aeronautics and Astronautics, Nanjing 211106, China (e-mail: {zhuqiuming}@nuaa.edu.cn).}
% \thanks{R. F. Gao is with the School of Transportation and Civil Engineering, Nantong University, Nantong 226019, China and also with the Nantong Research Institute for Advanced Communication Technologies, Nantong 226019, China
%  (e-mail: {grf}@ntu.edu.cn).}
% }

\maketitle
\thispagestyle{firstpage} % <--- 加上这一行，强制第一页使用我们定义的页眉

\begin{abstract}
\doublespacing

This study investigates the fundamental diffuse scattering mechanisms from three typical building wall surfaces, conducting measurements and model parameterization at 28 GHz and two key FR3 frequencies (8 GHz and 12 GHz). A novel three-dimensional (3D) measurement procedure is proposed to capture comprehensive spatial characteristics, and its effectiveness in improving parameterization accuracy was verified using 28 GHz data. For parameterization, we developed a new method utilizing two dimensions of the high-bandwidth power delay profile—received power and delay spread—thereby fully leveraging the rich information provided by such measurements. Furthermore, we introduce the ER-BK hybrid model, which integrates the Beckmann–Kirchhoff (BK) model's high accuracy and cross-frequency adaptability with the Effective Roughness (ER) model's simplicity, applying it to the building surfaces. Our results show that diffuse scattering at 8 GHz and 12 GHz is highly similar, distinct from that at 28 GHz. A comparison revealed that the BK model provides a better fit for our FR3 measurement data. Crucially, we validated the angular generalization of the parameterized BK model using data from a different incident angle than the one used for fitting. The feasibility of the ER-BK hybrid model was also verified through simulation of the parameterized marble surface.

% Through the research on the diffuse scattering behaviors of three typical outdoor surfaces at the three frequencies, we have proposed innovative methods in three aspects—measurement methods, parameterization methods, and models—for the study of diffuse scattering mechanisms in the FR3 band and mmWave frequencies, providing new ideas for subsequent diffuse scattering research.

\end{abstract}


\section{Introduction}
\label{sec:introduction}

\doublespacing
\IEEEPARstart{I}{n} future 6G networks, the FR2 band (24.25-71 GHz) and the FR3 band (7.125-24.25 GHz) has garnered significant attention from both academia and industry\cite{bazziupper}\cite{shakya2024comprehensive}. These two bands are particularly valued for their ability to address coverage, capacity, and deployment challenges in typical wireless scenarios. Additionally, they offer substantial advantages for emerging technologies, including non-terrestrial networks, reconfigurable intelligent surfaces, and integrated sensing and communications (ISAC)\cite {chen20235g}\cite {cui20236g}. 

% However, these two frequency bands operate at higher frequencies than sub-6 band. And it means relatively large path loss and significantly different propagation mechanisms.
% Therefore, the channel modeling for the FR2 and FR3 frequency bands needs to reconsider the propagation mechanisms of electromagnetic (EM) waves .

% Channel modeling traditionally employs two methodologies: deterministic and stochastic approaches. Stochastic methods reconstruct channels by leveraging empirical distributions and statistical properties derived from fitting measured data. In contrast, deterministic modeling, typically implemented through ray tracing, utilizes optical and electromagnetic propagation principles to develop radio channel models. While deterministic approaches eliminate the need for channel measurements, they require detailed environmental descriptions, including material parameters and propagation mechanisms.

Electromagnetic (EM) wave propagation mechanisms typically include free-space propagation, penetration, reflection, diffuse scattering, and diffraction\cite{han2014multi}. Among these, research on diffuse scattering is essential for reliable propagation analysis, particularly in the field of ISAC, as the modeling of scattered waves plays a critical role in sensing applications\cite{yaman2021ray}\cite{zhao2025sensing}\cite{reddy2020fundamental}. 
\textcolor{black}{In the sub-6 GHz band, diffuse scattering is negligible in most scenarios\cite{taleb2023transmission}\cite{ju2019scattering}. This holds true when the primary scatterers are building surfaces and relatively regular road surfaces, with minimal structural irregularities}
However, in the FR2 and FR3 bands, which operate at higher frequencies, diffuse scattering effects become significantly more pronounced

Diffuse scattering models have been extensively investigated through measurements and theoretical analysis. Full-wave simulations such as the finite-difference time-domain method, which is based on Maxwell's equations, have been used to study diffuse scattering models\cite{yi2023full}\cite{xie2022terahertz}. The physical optics approximation (PO) has also been applied in the research of diffuse scattering models\cite{hanpinitsak2019frequency}. However, both methods require detailed parameters of the surface and involve a high computational complexity. The works in \cite{1310631} and \cite{4052607} proposed integrating an ER based diffuse scattering model into ray tracing algorithms. This concept was subsequently validated through multitudinous measurements in the millimeter-wave band \cite{ren2020diffuse} \cite{pascual2016importance}. Another well-known DS model is the Beckmann-Kirchhoff (BK) model, which is based on the Kirchhoff approximation. It can describe the diffuse scattering properties across frequency bands using only two parameters: the root-mean-square (RMS) height of surface roughness and the spatial irregularity known as the correlation length\cite{beckmann1987scattering}. It performs well in the terahertz band\cite{das2023ambit}\cite{han2014multi}, however, its computational complexity is higher compared to the ER model.

Although existing studies have conducted measurements across multiple frequency bands and for various materials, there still exist a few limitations. First, to our best knowledge, few studies have explored diffuse scattering measurements in the FR3 band, and existing work lacks a robust integration of theoretical models and experimental data to evaluate the practical performance of diffuse scattering models in this frequency range. Second, most existing measurement efforts have focused only on experiments where the transmitter (Tx) and receiver (Rx) are in the same plane, resulting in a lack of acquisition of 3D spatial information of diffuse scattering. Third, existing parameterization methods only consider the angular spectrum of power, which fails to fully utilize high-bandwidth measurement data. Finally, the current ER model struggles with low accuracy and the BK model suffers from high complexity, a new approach is therefore needed to enhance diffuse scattering simulations. Addressing these limitations, our study makes four key contributions.

\begin{itemize}
\item We conduct measurements on three typical outdoor building surfaces at two representative frequencies in the FR3 band, namely 8 GHz and 12 GHz, along with the 28 GHz millimeter-wave frequency and collect high-bandwidth power delay profile (PDP) data.

\item We propose a 3D measurement procedure to capture richer spatial information of diffuse scattering and carry out real-world  measurements to verify the effectiveness of this procedure.

\item We propose a model parameterization method that combines the power angular spectrum and the delay spread angular spectrum, to make full use of the high-bandwidth PDP data.

\item We propose an ER-BK hybrid model which leverages the simplicity of the ER model and the cross-frequency band characteristics of the BK model. The effectiveness of this method has been verified using our measurement data.

\end{itemize}

% However, for the 7–24 GHz frequency band which lies between the aforementioned ranges—the diffuse scattering models developed for mmWave and terahertz applications may not retain their validity. Both theoretical frameworks and experimental measurements for diffuse scattering in this intermediate band remain underexplored, highlighting a critical research gap in the field.

% In this paper, we first introduce two important diffuse scattering models: the Effective Roughness (ER) model and the Beckmann-Kirchhoff (BK) model. We then present our scattering measurements conducted on three typical building wall surfaces at 8 GHz and 12 GHz. Leveraging the power and delay spread information from large-bandwidth power delay profile (PDP) data, we parameterize both models for the three materials and two frequencies using our in-house developed RT software. Finally, we analyze the implications of the parameterized BK model.

\section{Method}
\subsection{Diffuse scattering models}
\begin{figure}[!t]
    \centering   
    \includegraphics[width = 0.6\columnwidth]
    {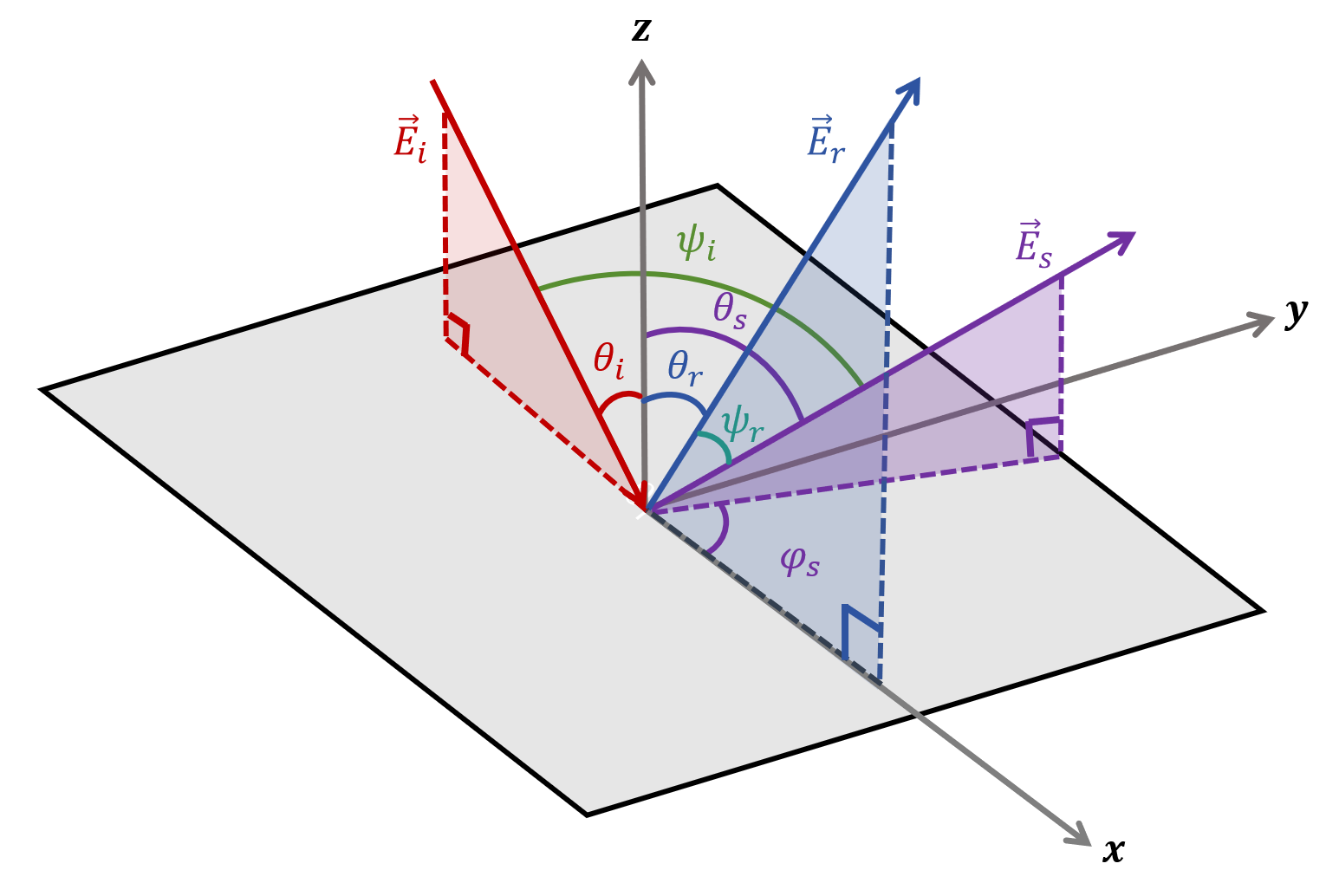}
    \caption{\textcolor{black}{Illustration of diffuse scattering geometry.}}   
    \label{scattergeo}
\end{figure}

\textcolor{black}{First, the general diffuse scattering geometry is shown in Fig. \ref{scattergeo}. The incident wave $\vec{E}_i$ has a zenith angle $\theta_i$ and an azimuth angle $\pi$. The reflected wave $\vec{E}_r$ features a zenith angle $\theta_r = \theta_i$ and an azimuth angle $0$, whereas the scattered wave $\vec{E}_s$ has a zenith angle $\theta_s$ and an azimuth angle $\phi_s$. Here, $\psi_i$ and $\psi_r$ denote the spatial angles between the scattered wave and incident wave directions, and between the scattered wave and reflected wave directions, respectively.}

\bibliographystyle{IEEEtran}

% \subsubsection{ER model}
The ER model, one of the most widely-used diffuse scattering models, characterizes the diffuse scattering process through a two-step approach.

First, it calculates the proportion of diffuse scattering energy to incident energy. Using the energy conservation relationship, the ratio of the scattered power to the incident power can be obtained as follow\cite{4052607}:

\begin{equation}
    S=\sqrt{(1-\rho^2)  \Gamma^2},
\end{equation}
\noindent where $\Gamma$ represents the smooth surface reflection
coefficient, and $\rho$ is the reflection reduction factor that can be estimated by
\cite{beckmann1987scattering}\cite{boithias1987radio}:
\begin{equation}
    \rho=\exp\left(-\frac{1}{2}\left(k^2 h_{\text{rms}}^2(\cos\theta_\mathrm{i}+\cos\theta_\mathrm{s})\right)^2\right),
\end{equation}
where $h_\mathrm{rms}$ represents  the standard deviation of the surface protuberance height about the mean height, \(k = {2\pi }/\lambda\) represents the wave number.

Second, energy-normalized diffuse scattering patterns can be described by empirical formulas, and there are several classic empirical formulas, including the Lambertian model, directive model, and backscattering lobe model.
 Each one represents a distinct physical scenario and accounts for different diffuse scattering patterns.

Directive model postulates that the primary energy concentration occurs along the specular reflection direction with a certain angular spread\cite{vitucci2023reciprocal}. The angular distribution of  the field intensity be  expressed as\cite{1310631} :

\begin{equation}
       \begin{array}{cc}
       {\left| {\bar{E}}_{s}\right| }^{2}\left( {{\theta }_{i},{\phi }_{i} = \pi ,{\theta }_{s},{\phi }_{s}}\right) \\=\left(\frac{K\cdot S}{r_i\cdot r_s}\right)^2\cdot\frac{\cos\theta_idS}{F_{\alpha_R}}\cdot\left(\frac{1+\cos\psi_R}{2}\right)^{\alpha_R}   
       \end{array}
       \label{3}
\end{equation}
Here, $K$ is a constant depending on the amplitude of the impinging wave:
\begin{equation}
     K = \sqrt{60 P_tG_t}  
     \label{K}
\end{equation}
Substituting it into equation (\ref{3}), we can get:
\begin{equation}
       \begin{array}{cc}
\left| \bar{E}_s \right|^2\left( \theta _i,\phi _i=\pi ,\theta _s,\phi _s \right) 
\\
=60\frac{\left| \bar{E}_i \right|}{240\pi}^2\frac{S^2}{{r_i}^2{r_s}^2}\frac{\cos \theta _i\mathrm{d}S}{F_{\alpha _R}}\cdot \left( \frac{1+\cos \psi _r}{2} \right) ^{\alpha _R}
\\
=\frac{\left| \bar{E}_i \right|^2S^2}{4{\pi r_i}^2{r_s}^2}\cdot \frac{\cos \theta _i\mathrm{d}S}{F_{\alpha _R}}\cdot \left( \frac{1+\cos \psi _r}{2} \right) ^{\alpha _R} 
       \end{array}
\end{equation}
where $r_i$ and $r_s$ denote the distances from the Tx antenna to the incident point and from the incident point to the Rx antenna, respectively. $\theta_i$ represent the incident angle. $\psi_s$ is the angle
between the diffuse scattering and the specular reflection directions. The parameter  $\alpha_R$  determines the beamwidth of the diffuse scattering pattern. As it decreases, the beamwidth expands, resulting in a more dispersed angular distribution of the scattered energy. $F_{\alpha R}$ is the normalization factor, and is given by :
\begin{equation}
        {F}_{{\alpha }_{R}} = \frac{1}{{2}^{{\alpha }_{R}}} \cdot  \mathop{\sum }\limits_{{j = 0}}^{{\alpha }_{R}}\left( \begin{matrix} {\alpha }_{R} \\  j \end{matrix}\right)  \cdot  {I}_{j}
\end{equation}
and
\begin{equation}
{I}_{j} = \frac{2\pi }{j + 1} \cdot  {\left\lbrack  \cos {\theta }_{i} \cdot  \mathop{\sum }\limits_{{w = 0}}^{\frac{j - 1}{2}}\left( \begin{matrix} {2w} \\  w \end{matrix}\right)  \cdot  \frac{{\sin }^{2w}{\theta }_{i}}{{2}^{2w}}\right\rbrack  }^{\left( \frac{1 - {\left( -1\right) }^{j}}{2}\right) }.
\end{equation}

The backscattering lobe model extends the directive model by incorporating an additional term to account for backscattering effects.  In practical scenarios where surfaces display significant irregularities (e.g., balconies or columns) and the incident angle approaches grazing incidence, diffuse scattering contributions become substantial near the incident direction. To address this, the model integrates a diffuse scattering lobe aligned with the incident direction. The  formulation is expressed as follows\cite{4052607}:

\begin{equation}
    \begin{split}
        |\bar{E}_S|^2(\theta_i, \phi_i = \pi, \theta_s, \phi_s) &= E_{S0}^2 \cdot \left\lbrack 
            \Lambda \left( \frac{1 + \cos \psi_r}{2} \right)^{\alpha_R} \right. \\
            &\left. + (1 - \Lambda) \left( \frac{1 + \cos \psi_i}{2} \right)^{\alpha_i}
        \right\rbrack \\
        \alpha_i, \alpha_R &= 1, 2, \ldots, N; \quad \Lambda \in [0, 1]
    \end{split}
\end{equation}

% In this model, the parameter $\alpha_i$ governs the width of the back lobe, while $\alpha_R$ controls the width of the forward lobe. As these parameters increase, the lobe width decreases, leading to more concentrated scattered energy. 
% In addition, $\Lambda$ is the distribution coefficient (with 
% $\Lambda\in[0,1]$
% ) that regulates the amplitude ratio between the forward lobe and backscattering lobe. A value of 
%  $\Lambda$ = 1
%  means the model reduces to the directive model with only the forward lobe contributing.
% $\psi_i$ denotes the angle between the scattered and incident directions, and $\psi_R$ denotes the angle between the scattered direction and specular reflection direction.

\textcolor{black}{In this model, $\psi_i$
 denotes the angle between the scattered and incident directions, and 
$\psi_R$
 denotes that between the scattered direction and specular reflection direction,
$\alpha_i$ governs the width of the back lobe, while 
 $\alpha_R$ controls that of the forward lobe—with larger values for both parameters narrowing their respective lobes and concentrating scattered energy. Additionally, 
 $\Lambda$ 
, a distribution coefficient within the range 
$\Lambda\in[0,1]$
, regulates the amplitude ratio between the forward and backscattering lobes; when 
 $\Lambda$ = 1
, the model simplifies to the directive model, retaining only the forward lobe contribution. }

The maximum amplitude $E_{S0}$ is calculated by\cite{1310631} :
\begin{equation}
    E_{S0}^2 = \left(\frac{K\cdot S}{r_i\cdot r_s}\right)^2\cdot\frac{\cos\theta_idS}{F_{\alpha_R,\alpha_i}}
\end{equation}
\noindent where
\begin{equation}
    F_{\alpha_R,\alpha_i}=\frac{\Lambda}{2^{\alpha_R}}\cdot\left[\sum_j^{\alpha_R}\begin{pmatrix}\alpha_R\\j\end{pmatrix}\cdot I_j\right]+\frac{(1-\Lambda)}{2^{\alpha_i}}\cdot\left[\sum_j^{\alpha_i}\begin{pmatrix}\alpha_i\\j\end{pmatrix}\cdot I_j\right]
\end{equation}
and 
\begin{equation}
    I_j=\frac{2\pi}{j+1}\cdot\left[\cos\theta_i\cdot\sum_{w=0}^{\frac{j-1}{2}}\begin{pmatrix}2w\\w\end{pmatrix}\cdot\frac{\sin^{2w}\theta_i}{2^{2w}}\right]^{\left(\frac{1-(-1)^j}{2}\right)}.
\end{equation}

% \subsubsection{BK Model}

Compared to the ER model, the two most critical parameters in the BK model are the variance of surface height $h_\mathrm{rms}$ and the surface correlation length $T$. The BK model assumes that the height distribution of the material surface follows a Gaussian profile\cite{{beckmann1987scattering}}\cite{tsang2000scattering}:
\begin{equation}
    W\left( {\Delta h}\right) = \left( {1/h_\mathrm{rms} \sqrt{2}}\right) \exp \left( {-\Delta {h}^{2}/2{h_\mathrm{rms}}^{2}}\right)
\end{equation}

\noindent and the surface correlation function is also Gaussian, satisfying:

\begin{equation}
    C\left( \tau \right) = \exp \left( {-{\tau }^{2}/{T}^{2}}\right)
\end{equation}

The primary distinction between the BK model and the ER model lies in the form of the diffuse scattering pattern. The diffuse scattering power distribution in the BK model can be expressed as\cite{beckmann1987scattering}\cite{tsang2000scattering}\cite{vernold1998modified}:

\begin{equation}
    \begin{split}
        |\bar{E}_S|^2(\theta_i, \phi_i = \pi, \theta_s, \phi_s) &= \frac{|\bar{E}_i|^2 \Gamma^2}{(4\pi)^2 r_i^2 r_s^2} \cdot \cos\theta_i \mathrm{d}S \cdot \left( \pi T^2 \right) \\
        &\cdot \frac{F^2(\theta_i, \theta_s, \phi_s)}{\exp\left[ g(\theta_i, \theta_s) \right]} \cdot \sum_{n=1}^{\infty} \left\{ \frac{g^n(\theta_i, \theta_s)}{n! \, n} \right. \\
        &\left. \cdot \exp\left[ \frac{-T^2}{4n} v_{xy}^2(\theta_i, \theta_s, \phi_s) \right] \right\},
    \end{split}
    \label{BK}
\end{equation}
where $\theta_i$ and $\theta_s$ represent the zenith angles of the incident field and the scattered field, respectively, as illustrated in Fig. \ref{scattergeo}. And

  \begin{equation}
      g\left( {{\theta }_{i},{\theta }_{s}}\right)  = {h_\mathrm{rms} }^{2}{v}_{z}^{2}\left( {{\theta }_{i},{\theta }_{s}}\right),
  \end{equation}
  \begin{equation}
      {v}_{x}\left( {{\theta }_{i},{\theta }_{s},{\phi }_{s}}\right) = k\left( {\sin{\theta }_{i} - \sin{\theta }_{s}\cos{\phi }_{s}}\right),
  \end{equation}
  \begin{equation}
      {v}_{y}\left( {{\theta }_{s},{\phi }_{s}}\right) = k\left( {\sin{\theta }_{s}\sin{\phi }_{s}}\right),
  \end{equation}
  \begin{equation}
    {v}_{z}\left( {{\theta }_{i},{\theta }_{s}}\right)  =- k\left( {\cos{\theta }_{i} + \cos{\theta }_{s}}\right),
\end{equation}
  Besides, F in (\ref{BK}) is called the geometric factor, which can significantly affect the diffuse scattering pattern. The most classic geometric factor is the Beckmann geometric factor\cite{ragheb2007modified}:
  \begin{equation}
{F}_{\text{ Beck }}\left( {{\theta }_{i},{\theta }_{s},{\phi }_{s}}\right)
= \frac{1 + \cos {\theta }_{i}\cos {\theta }_{s} - \sin {\theta }_{i}\sin {\theta }_{s}\cos {\phi }_{s}}{\cos {\theta }_{i}\left( {\cos {\theta }_{i} + \cos {\theta }_{s}}\right) }. 
\end{equation}
In addition, there is the Ogilvy factor, which is calculated based on boundary conditions:
\begin{equation}
    {F}_{\text{ Ogil }} = {F}_{\text{ Beck }}\cos \left( {\theta }_{i}\right)
\end{equation}

% \begin{figure}[!t]
%     \centering   
%     \includegraphics[width = 0.7\columnwidth]
%     {BK.png}
%     \caption{llustration of scattering pattern of BK model. Frequency
% of indicent wave is 30 GHz, $h_\mathrm{rms}$ = 2 mm, T = 1 cm}   
%     \label{scattergeo}
% \end{figure}
% \subsubsection{Comparation between ER model and BK model}
The ER model can simply describe the diffuse scattering effects of different types of surfaces. However, its empirical parameters, including $S$ and $\alpha$, are difficult to determine in practical applications. Even for the same surface, these parameters will change with the frequency of the incident wave and the incident angle. In contrast, the BK model computes scattering results across frequency bands and incident angles via physical calculations once its two parameters are determined. However, its algorithmic complexity remains a limitation.

It has been pointed out in \cite{hanpinitsak2025comparison}  that the ER model, such as the ER directive model, can be used to fit the BK model. 
Notably, the improved ER directive model is employed here, as the BK model's scattering results do not always exhibit peak energy in the specular reflection direction.
To achieve a better fit to the BK simulation results, we adjust the ER directive model, by replacing the original variable $\psi_s$ with a new variable $\psi_s'$, which represents the angle between the scattering direction and the peak scattering power direction.
This demonstrates that the ER model can accurately match the BK simulation results, significantly simplifying scattering calculations. Additionally, it enables efficient determination of cross-frequency scattering parameters in ray tracing.

\subsection{Measurement Campaign}

% \subsubsection{Measurment System }

We used a time domain channel sounding system based on National Instruments (NI) hardware to conduct the measurement campaign\cite{guo2025measurement}. The system operates under a superheterodyne architecture with an intermediate frequency (IF) range of 8 to 12 GHz. With the addition of our up-converter, it can also output millimeter waves spanning from 27.5 to 29.5 GHz.

The baseband signal of this system is a Zadoff-Chu (ZC) sequence with a length of 65,535. It employs an field programmable gate array module for real-time signal processing. Leveraging the autocorrelation properties of the ZC sequence, the PDP of the channel can be obtained\cite{sun2025modeling}\cite{10210347}, with an a multipath time delay resolution of up to 0.65 nanoseconds. Synchronization between the Tx and Rx is achieved using two pre-synchronized rubidium atomic clocks.

\begin{figure}[htbp] % [htbp] 指定图片位置优先级
    \centering % 整个图居中显示
    
    % --- 子图 (a) ---
    \begin{subfigure}[b]{0.7\textwidth} % [b] 表示底部对齐，宽度为版心的70%
        \centering
        \includegraphics[width=\textwidth]{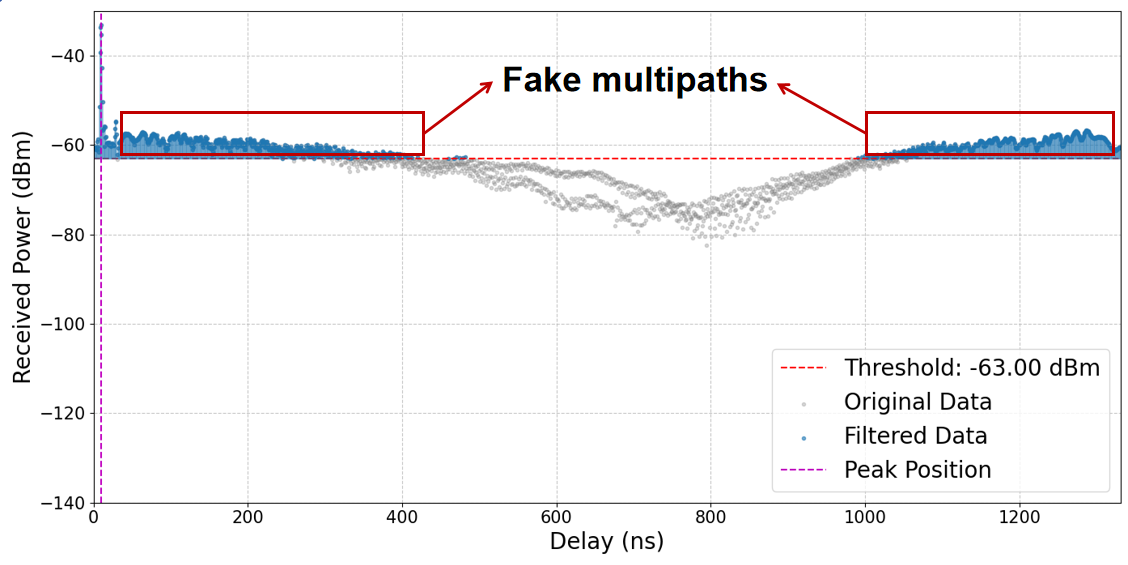}
        \caption{Typical PDP before calibration.} % 子图 (a) 的标题
        \label{fig:B2B} % 子图 (a) 的标签
    \end{subfigure}
    
    \vspace{10pt} % 两张子图之间的垂直间距
    
    % --- 子图 (b) ---
    \begin{subfigure}[b]{0.71\textwidth}
        \centering
        \includegraphics[width=\textwidth]{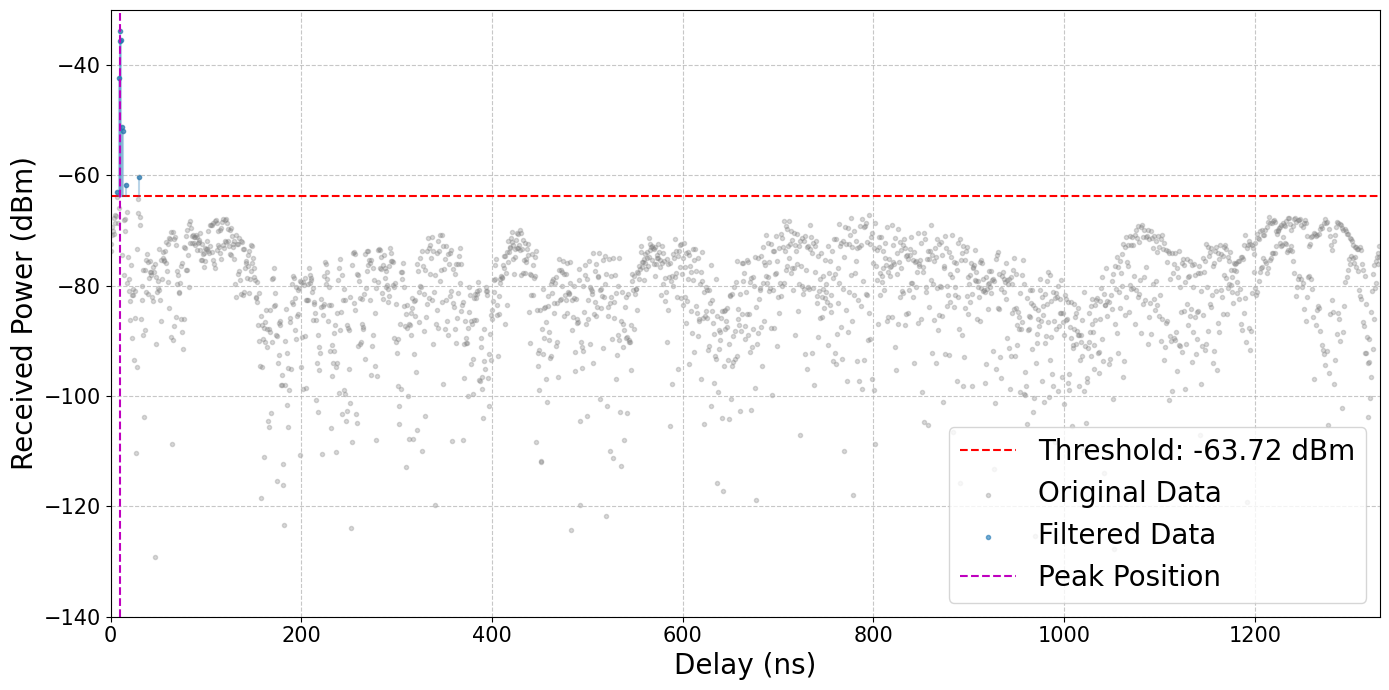}
        \caption{Typical PDP after calibration.} % 子图 (b) 的标题
        \label{fig:AFB2B} % 子图 (b) 的标签
    \end{subfigure}
    
    \caption{Comparison of PDP before and after calibration.} % 整个图的总标题
    \label{c} % 整个图的总标签
\end{figure}

% \subsubsection{Calibration \textcolor{blue}{ and Filtering}}
% \begin{figure}[htbp]
%   \centering
%   \begin{minipage}[t]{0.8\linewidth}  % 这里图片位置设置为[t]竖直优先
%   % {0.5\linewidth} 图片是页面高度的0.5倍
%       \centering
%       \label{BFB2B}\includegraphics[width=3.2in]{B2B.png}
%       % [width=2in] 图片宽度设置为2英寸，这里也可以用厘米
%       \caption{Raw PDP}
%   \end{minipage}
%   \begin{minipage}[t]{0.8\linewidth}
%       \centering
%       \label{AFB2B}\includegraphics[width=3.2in]{AFB2B.png}
%       \caption{PDP after B2B Calibration}
%   \end{minipage}
%   \caption{B2B Calibration}
%   \label{calibration}
% \end{figure}
The PDP data directly obtained from the channel sounding system includes the hardware response of the system. When using such data, \textcolor{black}{employing the  widely accepted threshold (peak value of the PDP minus 30 dB) to get the filtered PDP data results in  numerous fake multipath components that do not exist in reality}  (such as the multipaths appearing hundreds of nanoseconds after the peak in Fig. \ref{c} (a), corresponding to path lengths of tens to hundreds of meters, which do not exist in actual diffuse scattering meassurement with path of hundreds of meters).

To eliminate the hardware response, we employed a back-to-back calibration\cite{10210347}:
\begin{equation}
    {h}_{\mathrm{{cal}}}\left\lbrack  n\right\rbrack   = \operatorname{IFFT}\left( {{H}_{\mathrm{{raw}}}\left\lbrack  k\right\rbrack  /G\left\lbrack  k\right\rbrack  }\right)
\end{equation}
where $ h_{\text{cal}} $ denotes the PDP after calibration, while $ h_{\text{raw}} $ represents the raw PDP. The term $ G[K] $ is derived by directly connecting the Tx and Rx using a cable.
After calibration, as Fig. \ref{c} (b) shows, the previously observed fake multipaths are shown to have disappeared, with a significant reduction in the noise floor. 

\begin{figure}[htbp]
    \centering
    \subfloat[marble]{% 子图1标题
        \includegraphics[width=0.3\textwidth]{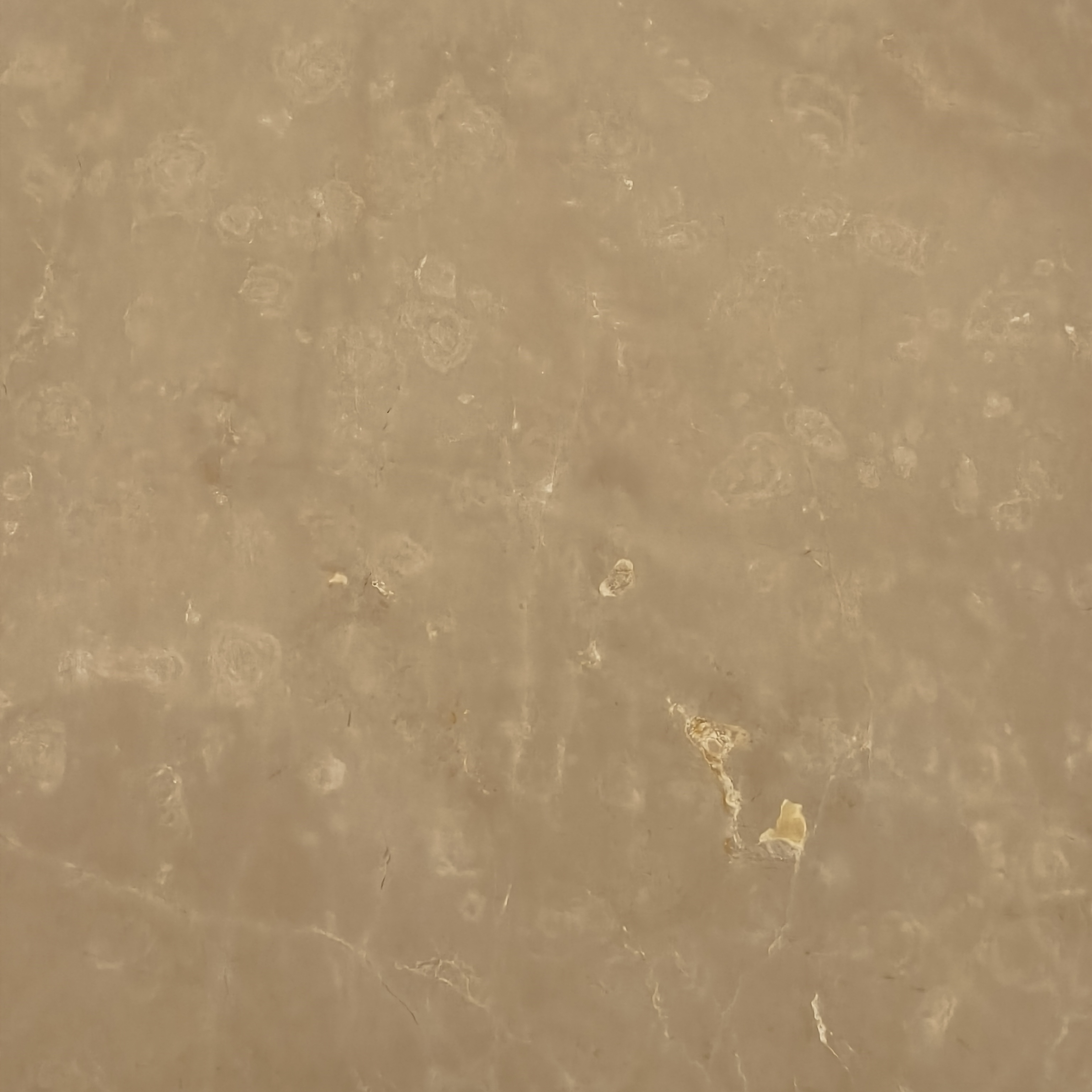} % 极坐标图路径
        \label{marble}
    }
    \subfloat[smooth wall]{% 子图2标题
        \includegraphics[width=0.3\textwidth]{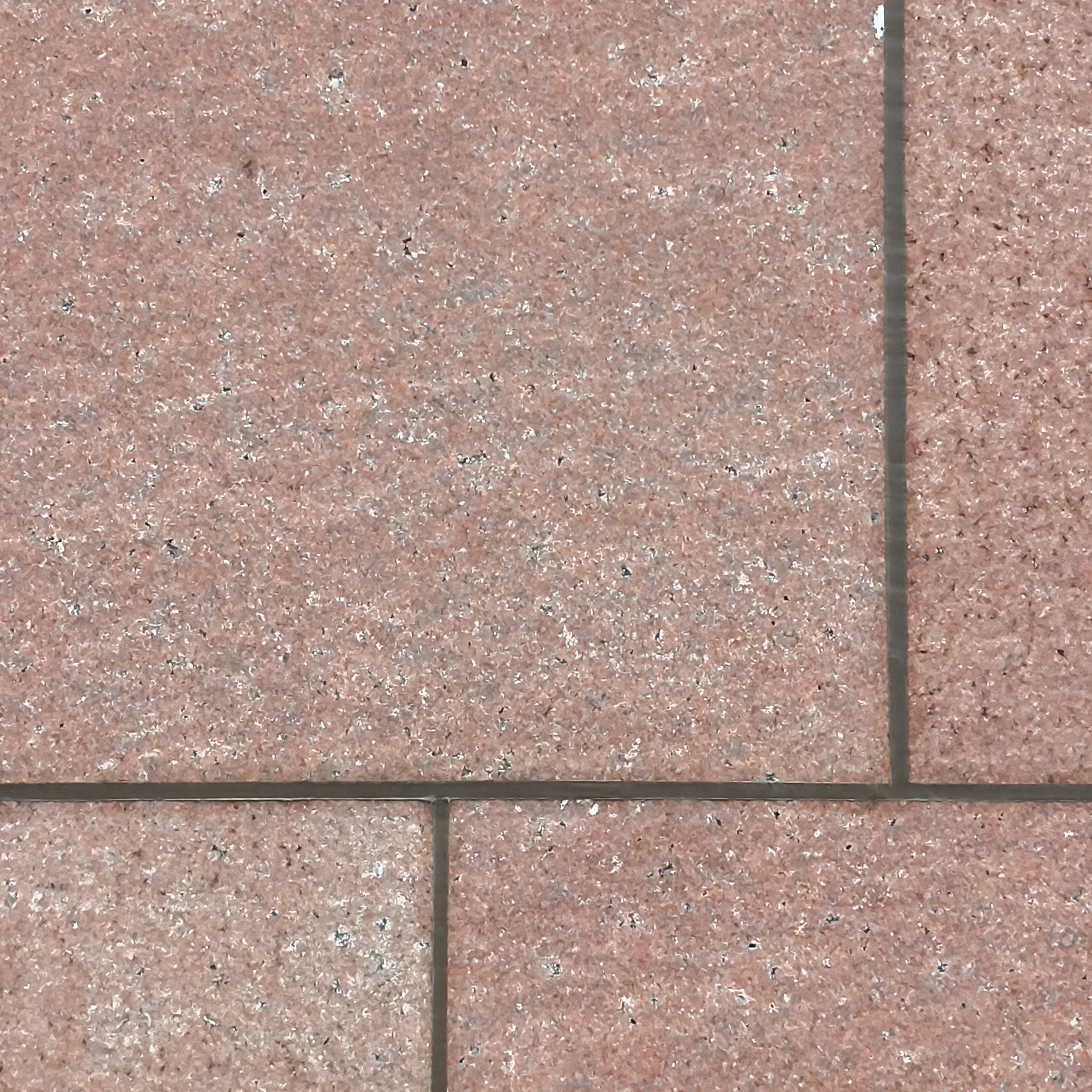} % 直角坐标图路径
    \label{wenbo}
    }
    \subfloat[rough wall]{% 子图2标题
        \includegraphics[width=0.3\textwidth]{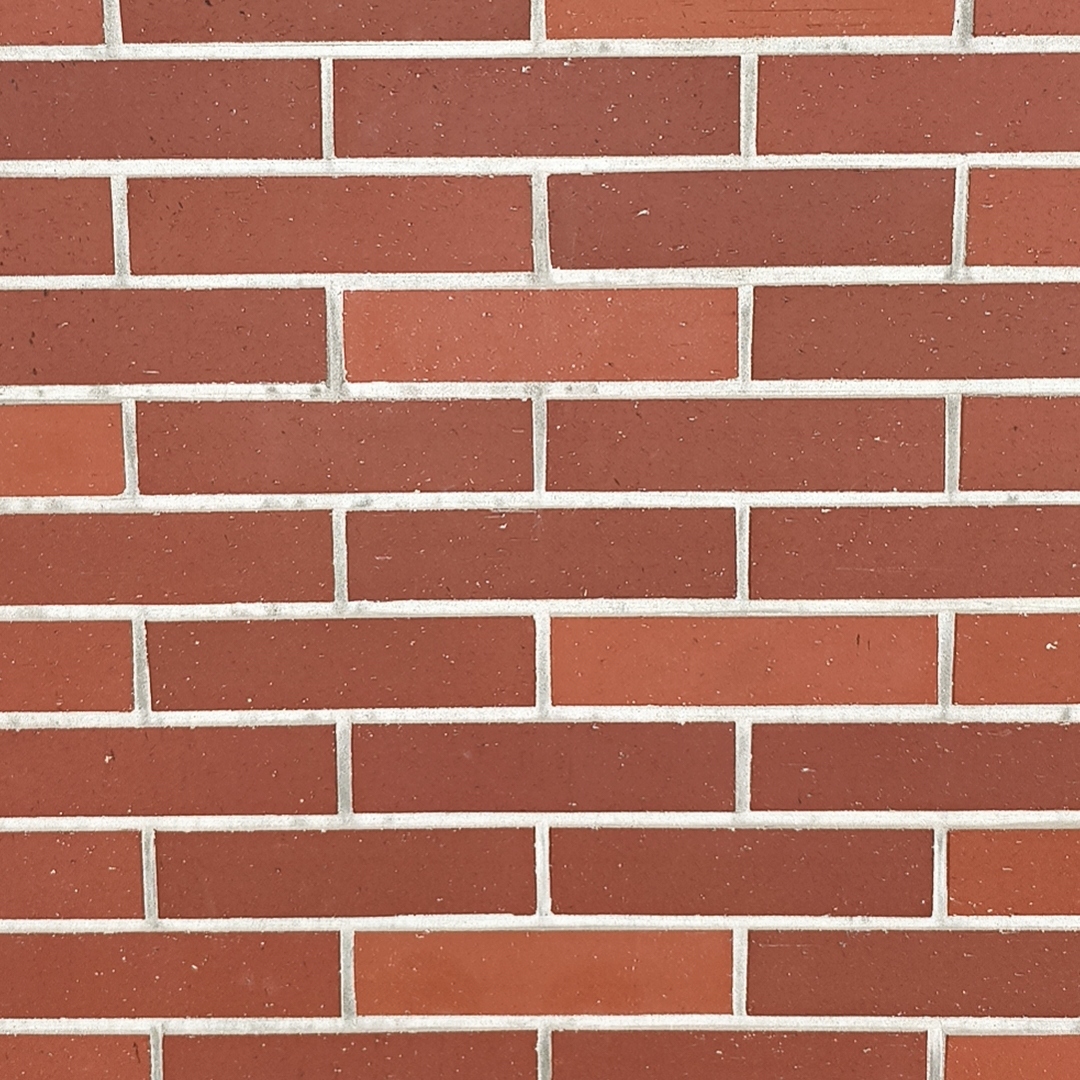} % 直角坐标图路径
        \label{brick}
    }
    \caption{Three building surfaces measured.} % 主标题
\label{material}
\end{figure}

\textcolor{black}{After calibration, noise at the several-hundred-nanosecond mark may still appear above the threshold. To mitigate this risk, we have therefore added a 20 ns window around the peak, a setting that captures all multipaths within 6 meters while filtering out unreasonable noise points.}
We conducted measurement campaigns on three types of building surfaces at the Minhang campus of Shanghai Jiao Tong University at 8 GHz, 12 GHz , 28 GHz. The surfaces of the measured building walls are shown in Fig.\ref{material}.

The measurements used two tripod-mounted horn antennas at the Tx and Rx, respectively. Their technical specifications at 8 GHz, 12 GHz and 28 GHz frequencies are detailed in Table \ref{table1}.
Four types of measurements were designed and conducted on the aforementioned three materials at the three carrier frequencies.

\begin{figure}[!t]
    \centering
    \includegraphics[width=0.8\columnwidth]
    {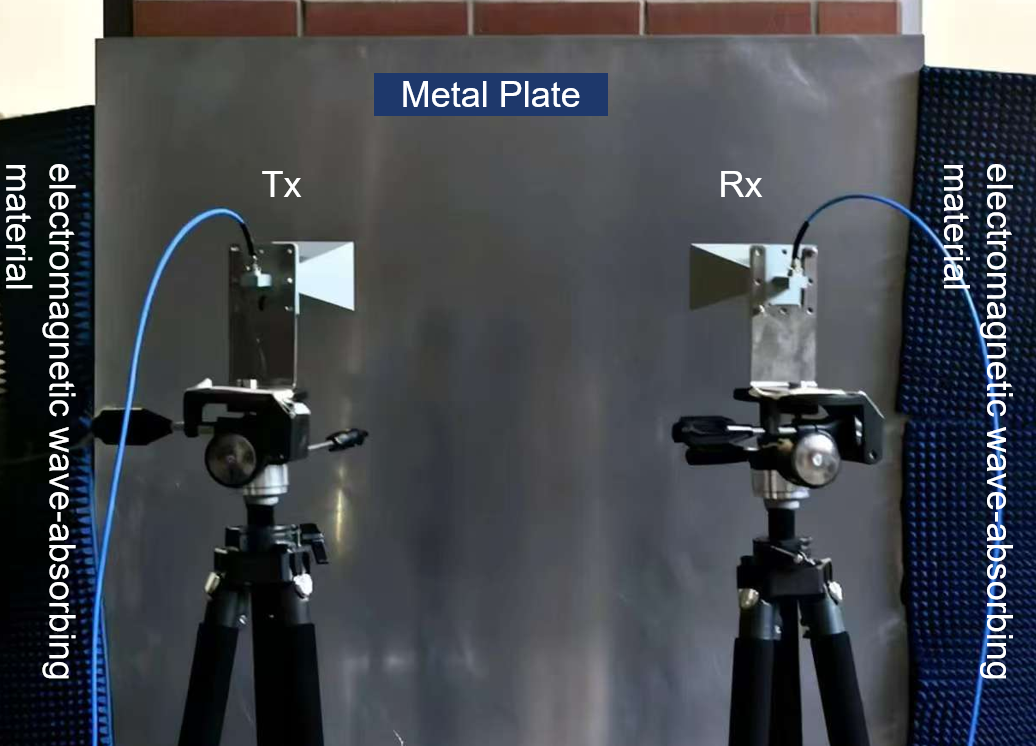}
    \caption{\textcolor{black}{Measurement scenario of metal plate reflection.}}      
    \label{sen}
\end{figure}

\begin{figure*}[!t]
    \centering
    \begin{subfigure}{0.45\textwidth}
        \centering
        \includegraphics[width=\linewidth]{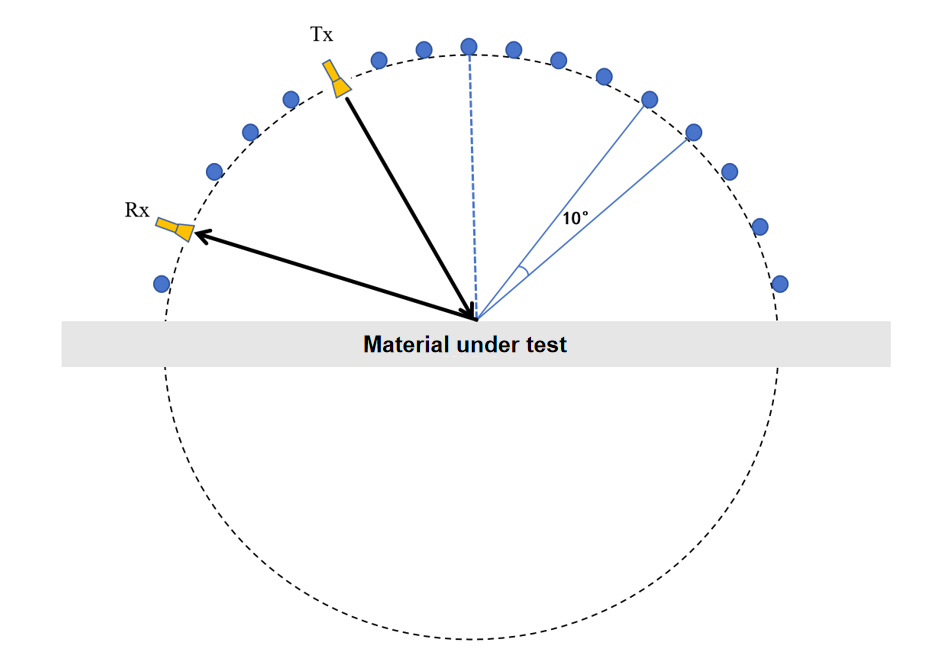}
        \caption{Illustration of measurement setup in a plane.}
    \end{subfigure}%
    \hfill
    \begin{subfigure}{0.45\textwidth}
        \centering
        \includegraphics[width=\linewidth]{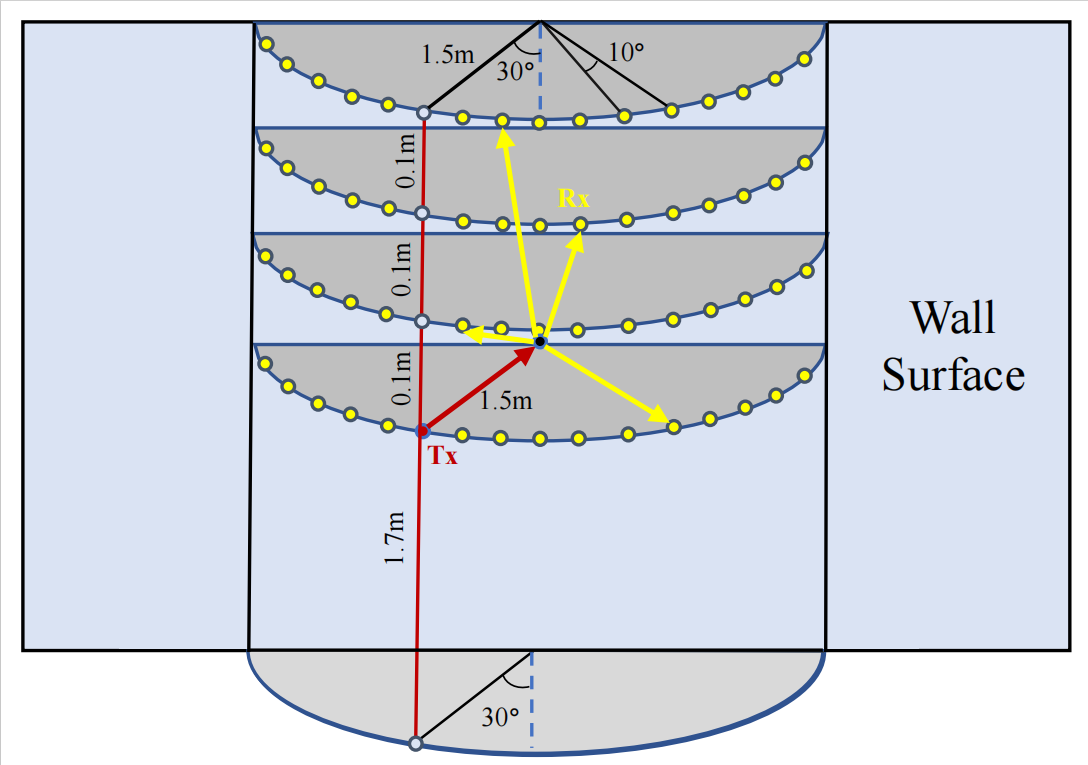}
        \caption{Illustration of measurement setup in a 3D scenario.}
   
    \end{subfigure}
    \caption{Illustration of measurement setup.}
    \label{measen}
\end{figure*}

\begin{itemize}

    \item \textit{Tx and Rx antennas at the same height, incidence angle $= 30$°:}

Fig. \ref{measen} (a) depicts the measurement configuration with the Tx and Rx in the same plane, in which the Tx antenna is fixed at a height of 1.7 meters and positioned 1.5 meters from the center of the target wall surface \textcolor{black}{( which is significantly greater than the Rayleigh distances of the antenna corresponding to each of the three frequencies, thereby ensuring the far-field condition is met for all test cases)}, oriented at a 30-degree angle relative to the normal direction. The Rx antenna is placed at the same height and distance from the center, and is systematically rotated in 10-degree increments. 

    \item \textit{Tx and Rx antennas at the same height, incidence angle $\neq30$°:}
\textcolor{black}{To evaluate the generalization ability of our parameterized model under other incidence angles for the same surface, we performed 
measurements on marble and smooth wall surfaces with Tx angles of 40°, 50°, and 60°, and Rx angles fixed at 0° and the specular reflection direction, respectively.}

    \item \textit{Metal plate reflection measurement:}
\textcolor{black}{
To verify the rationality of our measurement setup and simulation method, we conducted measurements on the power-angle spectrum of metal plate reflections at the three frequencies, as shown in Fig. \ref{sen}.  
Specifically, using the same measurement setup as illustrated in Fig.\ref{measen} (a) — with an incidence angle of 30°, the Rx rotated in 10-degree increments, and electromagnetic wave absorbing materials placed on both sides to prevent scattering interference from areas other than the target wall — the only difference is that the target wall surface was replaced with a metal plate.}

\item \textit{Tx and Rx antennas at different heights:}

In addition, to provide more information for model parameterization from the measured data, as suggested in \cite{guo2025measurement}, additional spatial angle measurement can be implemented. As Fig. \ref{measen} (b), we conduct the spatial angle measurements on the rough wall. For the measurement, the Tx antenna is fixed at a height of 1.7 meters in the horizontal plane, aligned with the center of the wall and at a 30-degree angle to the normal. The Rx antenna is adjusted to heights of 1.7 meters, 1.8 meters, 1.9 meters, and 2.0 meters, while moving along a circular path with a radius of 1.5 meters, with a step size of 10 degrees for each measurement.
\end{itemize}
 
% \begin{table}[!]
% \centering
% \caption{\textsc{Parameters of Experiment Setup}}
% \label{parameter_table}
% \begin{tabular}{cccccc}
% \toprule
% Frequency  &   Gain (dBi)& HPWB (°)& Transmission Power (dBm)\\ \midrule
% 8GHz &      19.4   & 18.3  & 10\\
% 12GHz &      21.8   & 12.5  & 10\\ \bottomrule
% \end{tabular}
% \label{table1}
% \end{table}

\begin{table}[htbp]
    \centering
    \caption{\textsc{Parameters of Experiment Setup}}
    \renewcommand{\arraystretch}{2.2}
    \begin{tabular}{|l|c|c|c|}
        \hline
        \textbf{Parameters} & \multicolumn{3}{c|}{\textbf{Frequency}} \\ \cline{2-4}
                           & 8 GHz & 12 GHz & 28 GHz \\ \hline
        Bandwidth (GHz)          & \multicolumn{3}{c|}{1.532} \\ \hline  % 三列合并
        Transmit Power (dBm)     & \multicolumn{3}{c|}{10} \\ \hline       % 三列合并
        Baseband Signal          & \multicolumn{3}{c|}{ZC Sequence} \\ \hline  % 三列合并
        Multipath Delay Resolution (ps) & \multicolumn{3}{c|}{650} \\ \hline  % 三列合并
        Antenna Type             & \multicolumn{3}{c|}{Horn} \\ \hline  % 三列合并
        Antenna Gain (dBi)       & 19.4 & 21.8 & 15 \\ \hline
        Antenna HPBW (°)         & 18.7 & 12.5 & 23 \\ \hline
    \end{tabular}
    \label{table1}
\end{table}
\subsection{Simulation and Parameterization}

% \subsubsection{Simulation}
We conducted simulations of the measurement scenario based on the ray-tracing algorithm. To enable the integration of different diffuse scattering models, we utilized a self-developed ray-tracing software. \textcolor{black}{Parameters involed in ray tracing such as the antenna's gain and HPBW are its actual parameters, which are provided in Table \ref{table1}.
}

\textcolor{black}{
In the simulations, the multi-hop paths and the ray paths from ground reflection and diffuse scattering have minimal impact on the PDP. To demonstrate this, we compared the measured reflection power spectra of the metal plate at three frequencies with the simplified ray-tracing simulated power angle spectrum (considering only the specular reflection of the metal plate) and calculated the RMSE between them. The RMSE values for 8 GHz, 12 GHz, and 28 GHz are 5.41 dB, 4.89 dB, and 4.74 dB, respectively. Such small discrepancies between the simplified simulation results and the measured data indicate that, under our measurement setup, the energy of the PDP is primarily derived from first-order scattering and reflection on the target surface.}

% \subsubsection{Parameterization Method}

We parameterize the diffuse scattering model by minimizing the discrepancy between measured data and ray-tracing simulation results. The high-bandwidth PDP data, featuring a time resolution of 0.65 ns (equivalent to a multipath resolution of approximately 0.195 m), enables enhanced multipath characterization. Consequently, our parameterization process extends beyond conventional power angular spectrum analysis\cite{pascual2016importance}\cite{ju2019scattering}, providing a more comprehensive framework than prior scattering model studies.

We integrate both the angular spectrum of the delay spread and the angular spectrum of power derived from the PDP of diffuse scattering measurement to fit the diffuse scattering model:
\begin{equation}
\frac{\sum_{k=1}^N{\left( l\left( \tau _{\mathrm{RMS}_k},\hat{\tau}_{\mathrm{RMS}_k} \right) +l\left( \mathrm{P}_k,\mathrm{\hat{P}}_k \right) \right)}}{2N}
\end{equation}
where:
\begin{equation}
    l\left(x,y\right)=\frac{\left|x-y\right|}{x+y}\in\left[0,1\right]
\end{equation}
The metric employed is the symmetric mean absolute percentage error (SMAPE), an accuracy measure based on percentage errors with both lower and upper bounds; a smaller value indicates higher model accuracy. It is widely used in assessing the performance of channel models
 \cite{hoydis2024learning}\cite{10971195}\cite{chen2024similarity}.
 $\mathrm{P}_{k}$ and $\tau _{\mathrm{RMS}_{k}}$ represent the received power and delay spread at each location in the measurement, while $\mathrm{\hat{P}}_{k}$ and $\hat{\tau}_{\mathrm{RMS}_{k}}$ denote the received power and delay spread at the corresponding locations in the simulated scenario.\textcolor{black}{N
 stands for the number of measured locations; for example, in measurements where the Tx and Rx are in the same plane, N = 16.
A key advantage of SMAPE is that it allows the fitting accuracy of different parameters (i.e. power and delay spread) to be compared collectively.} 
% \subsubsection{ER-BK hybrid model}
Combining the simplicity of the ER model and the accuracy of the BK model in prediction across frequency bands and under arbitrary incident angles, we propose the ER-BK hybrid diffuse scattering model.

\begin{itemize}
    \item \textit{Get parameters for BK model ($T$ and $h_{\text{rms}}$):} First, we acquire the two parameters $T$ and $h_{\text{rms}}$ of the surface in accordance with the measurement and parameterization scheme we proposed above.
    \item  \textit{Simulation based on BK model:} Second, based on these two acquired parameters, we perform simulations of the BK model for the target frequency to obtain diffuse scattering patterns at different angles within this frequency band.
    \item  \textit{Fit simulation result with ER model:} Third, we use the modified ER Directive model to fit the simulation results of the BK model, so as to obtain the three empirical parameters $S$, $\alpha$, and $\theta_p$ (angle of the peak of the diffuse scattering power) of the ER Directive model under the desired frequency and different incident angles.
\end{itemize}
% \subsubsection{Get parameters for BK model: $T$ and $h_{\text{rms}}$}
% First, we acquire the two parameters $T$ and $h_{\text{rms}}$ of the surface in accordance with the measurement and parameterization scheme we proposed above.
% \subsubsection{Simulation based on BK model}
% Second, based on these two acquired parameters, we perform simulations of the BK model for the target frequency to obtain scattering patterns at different angles within this frequency band.

% \subsubsection{Fit simulation result with ER model}
% Third, we use the modified ER Directive model to fit the simulation results of the BK model, so as to obtain the three empirical parameters $S$, $\alpha$, and $\theta_p$ (angle of the peak of the scattering power) of the ER Directive model under the desired frequency and different incident angles.

\begin{figure}[!t]
    \centering   
    \includegraphics[width = 0.8\columnwidth]
    {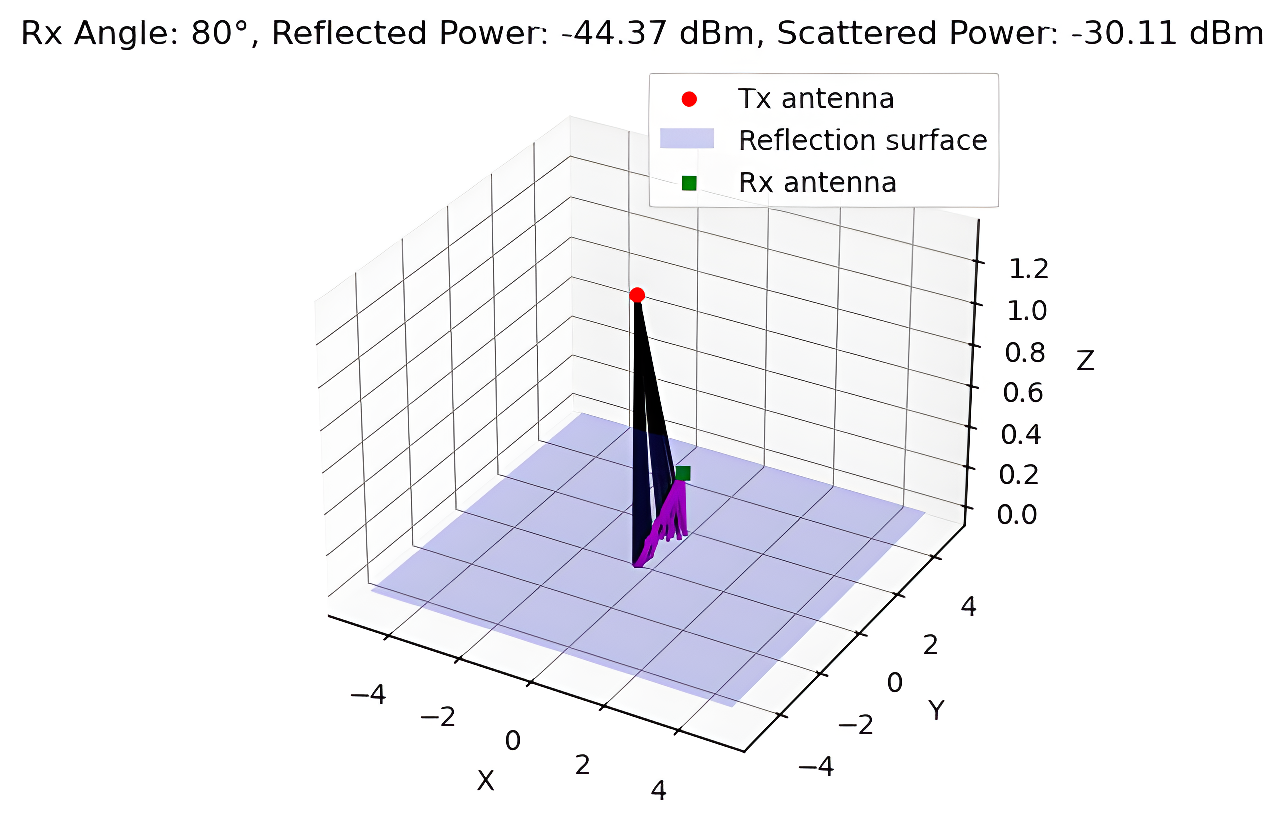}
    \caption{\textcolor{black}{Illustration of simulation scenario.}}   
    \label{sim_sen}
\end{figure}

\section{Results}
\subsection{Effectiveness of 3D Rx Measurements for Parameterization of DS Models}
As an example, we use the 3D Rx measurement data to parameterize the 28 GHz backscattering lobe model. In traditional parameterization processes \cite{pascual2016importance}\cite{guo2024diffuse}\cite{tian2019effect}, only the RT results within the incident plane are considered for fitting the measured data to determine the diffuse scattering model parameters, as demonstrated previously. This fitting process overlooks the changes caused by the spatial distribution of diffuse scattering power. To address this, under rough wall measurement conditions, we varied the height of Rx by $\Delta h$, integrated the obtained data, and then applied the fitting process described above. 

\begin{figure}[!t]
    \centering
    \includegraphics[width=0.8\columnwidth]
    {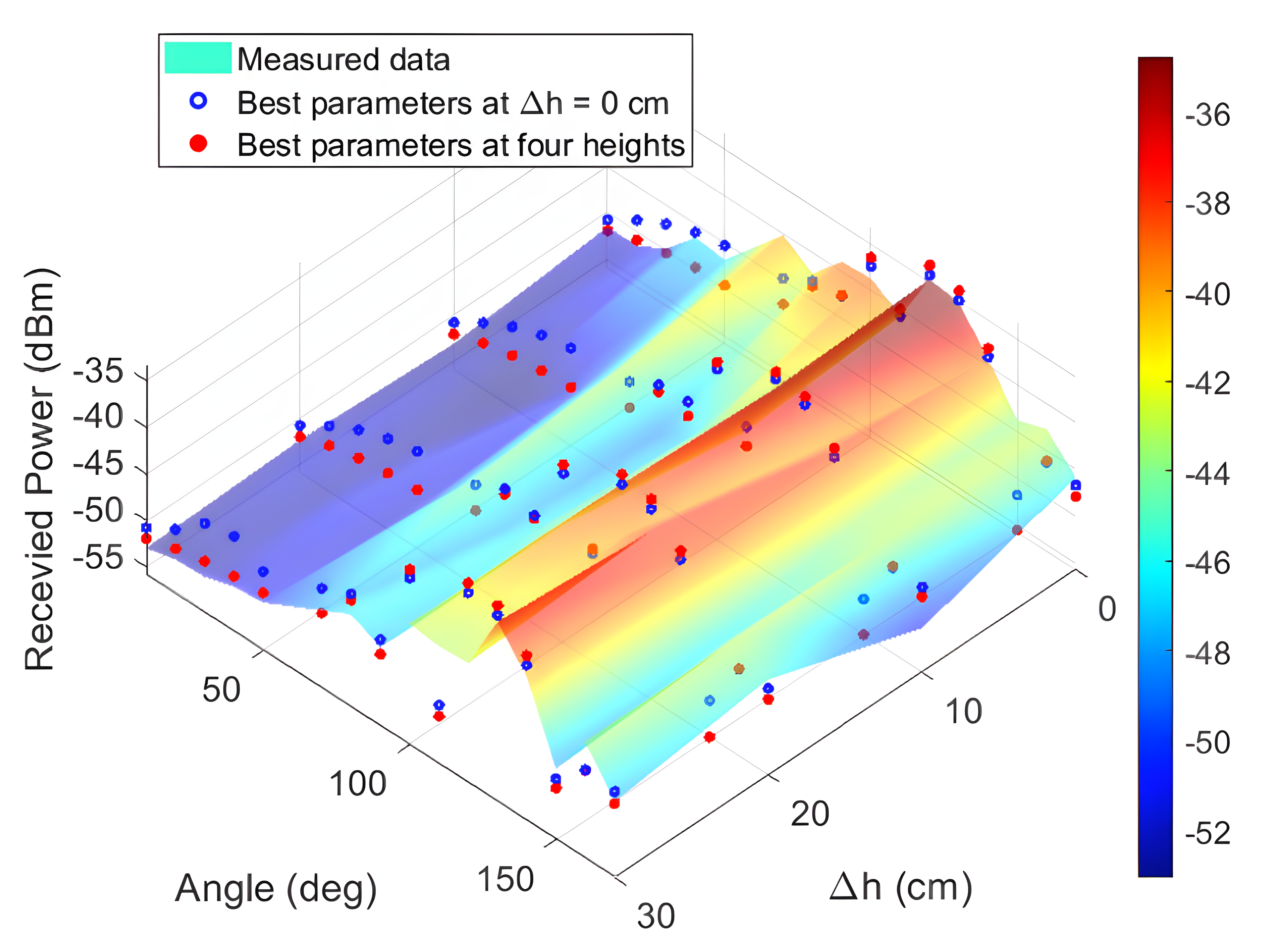}
    \caption{\textcolor{black}{Recevied power of different positions.}}      
    \label{fig6}
\end{figure}
The received power at various 3D positions is illustrated in Fig. \ref{fig6}, where the blue circles represent the results of the ray tracing with the best parameters using data only in the incident plane ($\Delta h=0$ cm), the red dots represent the results of the ray tracing with the best parameters at four Rx heights ($\Delta h=0, 10, 20, 30$ cm), and the colored surface represents the measured data.
As the Rx height increases, it can be observed that the diffuse scattering model determined based on the incident plane ($S$ = 0.60, $\alpha_R=1$, $\alpha_i=10$, $\Lambda=0.2$) incorrectly predicts the actual backscattering power outside the plane. On the other hand, by fitting the data obtained from multiple stereoscopic positions, the scattering model ($S$ = 0.42, $\alpha_R=6$, $\alpha_i=4$, $\Lambda=0.2$) can partially correct this error while still maintaining a good fitting performance within the incident plane. This indicates that for a more accurate and detailed analysis of diffuse scattering effects, it is necessary to consider the  diffuse scattering effects in the stereoscopic space, within an acceptable level of complexity, to jointly determine the actual diffuse scattering model parameters.

\begin{table*}[!ht]
    \centering
    \caption{Fitting parameters of BK and ER models for different materials}
    \label{model_fitting}
    \setlength{\tabcolsep}{12pt} % 增加列间距
    \begin{tabular}{ccccccccc}        \toprule
        Material & Model & \multicolumn{6}{c}{Fitting Parameters} & SMAPE \\
        \cmidrule(lr){3-8}
        & &  $\epsilon_r$ & $h_{\text{rms}}$ (mm) &  $\alpha_R$ & $\alpha_i$ & $\Lambda$ & T (mm) \\
        \midrule
        Marble Wall & ER Directive model & 6.1 & 1.1 & 1 & - & - & - & 0.3265 \\
                  & BK model & 6.2  & 1.0 & - & - & - & 5.0 & 0.2905 \\
        \addlinespace
        Smooth Wall & ER Directive model & 6.0 & 4.2 & 3 & - & - & - & 0.3514 \\
                  & BK model & 5.7 & 4.1 & - & - & - & 0.8 & 0.2833 \\
        \addlinespace
        Brick Wall & Backscattering lobe model & 10.1 & 8 & 1 & 4 & 0.8 & - & 0.3008 \\
                 & BK model & 11.5 & 6.5 & - & - & - & 2.1 & 0.2168 \\
        \bottomrule
    \end{tabular}
\end{table*}

\subsection{Material and Frequency Dependence of Diffuse Scattering Effect Among 8 GHz, 12 GHz and 28 GHz}
% Fig. \ref{frequency} (a)-(f) displays the angular power spectra and delay spread spectra for three material surfaces at two frequencies. As shown in Fig. \ref{frequency} (a)-(c), the received power for all three materials at both frequencies is concentrated in the specular scattering direction, exhibiting a certain degree of angular broadening. Fig. \ref{frequency} (d)-(f) demonstrate that the angular spread of the PDP is more pronounced within the incident plane. Furthermore, by comparing the angular power spectra and delay spread spectra across different materials and frequencies, we observe that the change from 8GHz to 12GHz has minimal impact on the scattering power and delay spread, while the material surface significantly affects both. Therefore, in the subsequent discussion on the fitting performance of the scattering model, we only present the fitting results from the 8GHz scattering measurement.

\begin{figure*}[htbp]
    \centering
    \begin{subfigure}{0.42\textwidth}
        \centering
        \includegraphics[width=\linewidth]{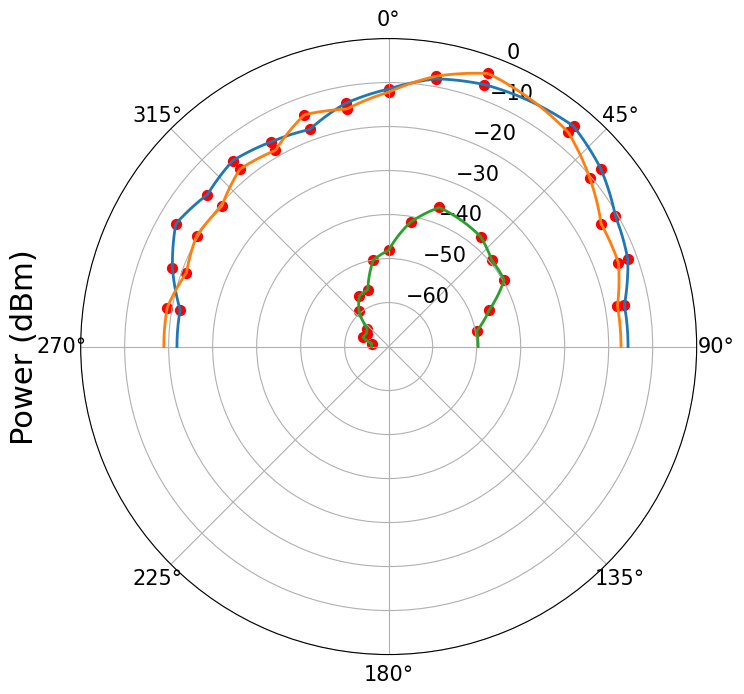}
        \caption{Marble wall power.}

    \end{subfigure}%
    \hfill
        \begin{subfigure}{0.42\textwidth}
        \centering
        \includegraphics[width=\linewidth]{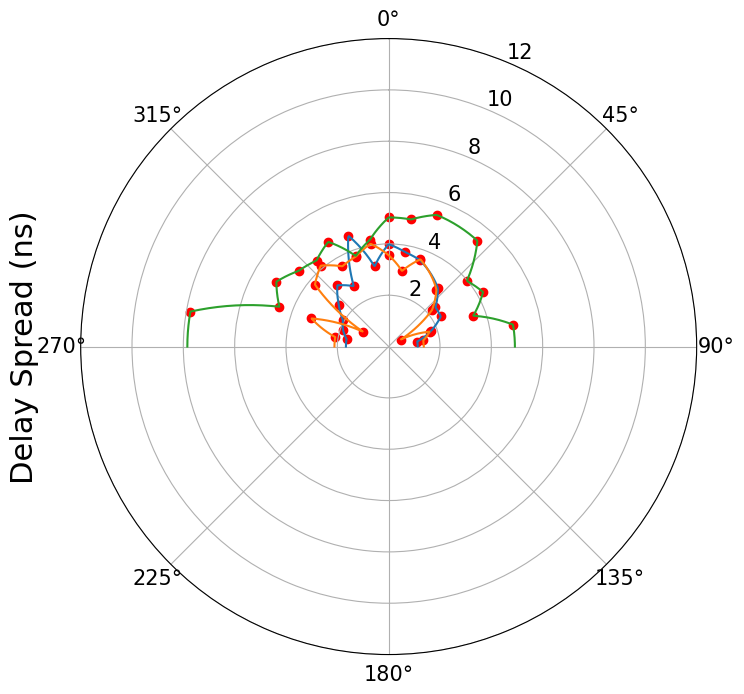}
        \caption{Marble wall delay spread.}

    \end{subfigure}%
    \hfill
            \begin{subfigure}{0.42\textwidth}
        \centering
        \includegraphics[width=\linewidth]{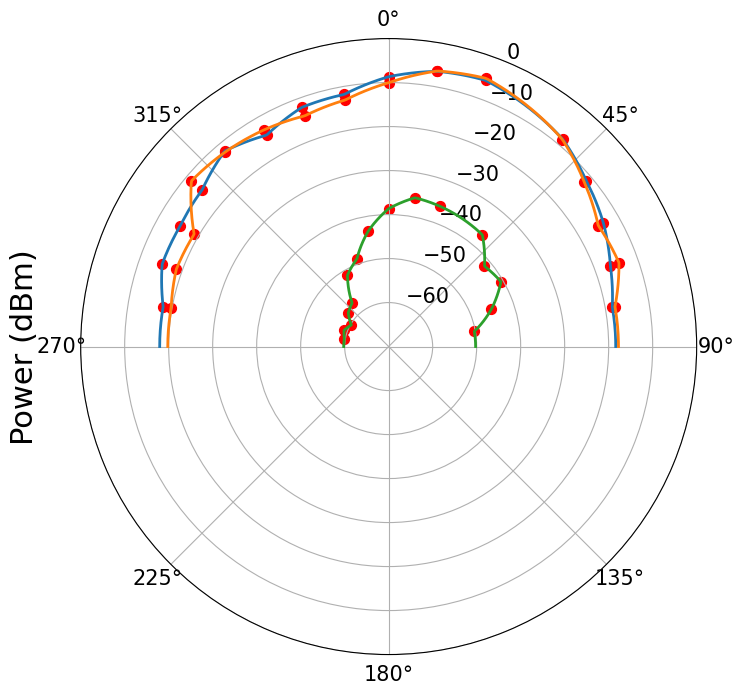}
        \caption{Smooth wall power.}
   
    \end{subfigure}
    \hfill
        \begin{subfigure}{0.42\textwidth}
        \centering
        \includegraphics[width=\linewidth]{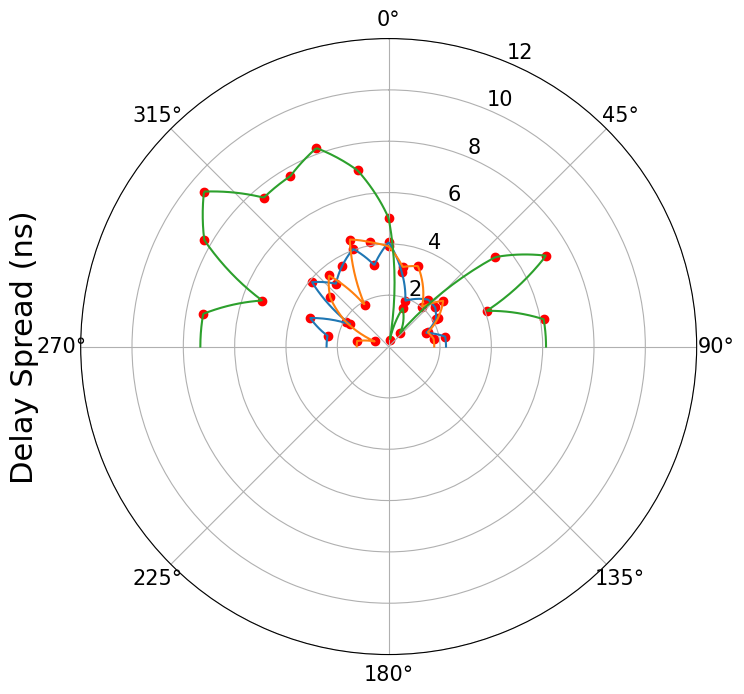}
        \caption{Smooth wall delay spread.}
  
    \end{subfigure}%
    \hfill
    \begin{subfigure}{0.42\textwidth}
        \centering
        \includegraphics[width=\linewidth]{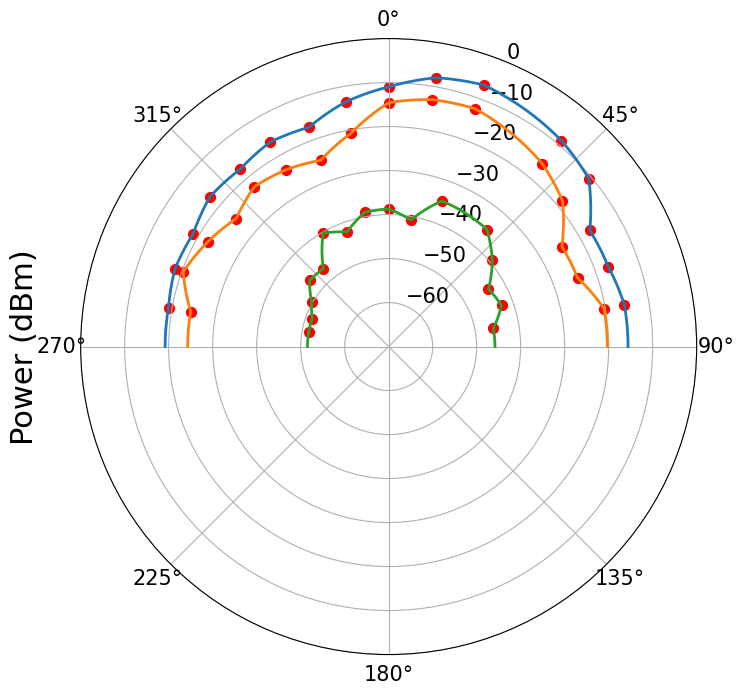}
        \caption{Brick wall power.}

    \end{subfigure}%
    \hfill
    \begin{subfigure}{0.42\textwidth}
        \centering
        \includegraphics[width=\linewidth]{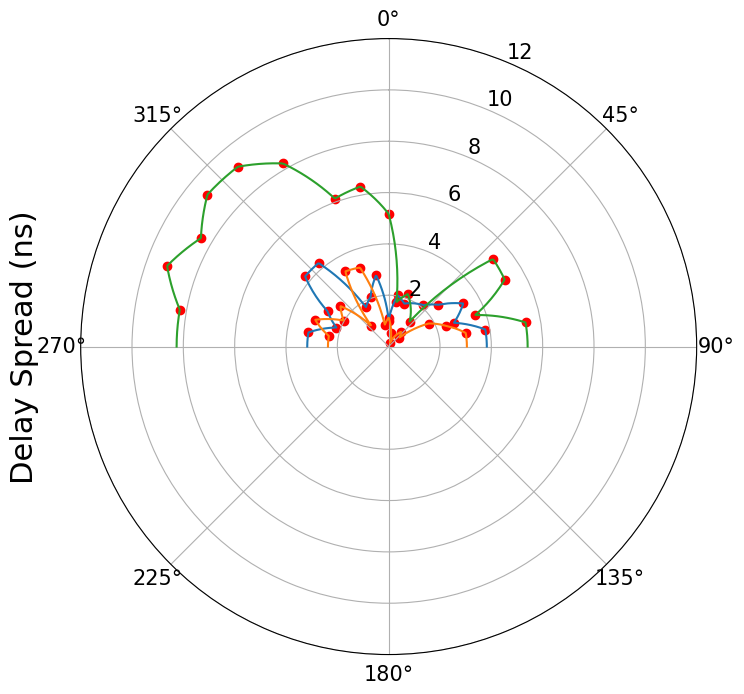}
        \caption{Brick wall delay spread.}
 
    \end{subfigure}
    \caption{(a), (c), (e) Angular spectrum of received power for three different surfaces at  8 GHz, 12 GHz and 28 GHz, (b), (d), (f) Angular spectrum of the delay spread of the PDP for these surfaces at the same three  frequency. Blue, orange and green perspectively represent 8 GHz, 12 GHz and 28 GHz.  }
    \label{frequency}
\end{figure*}

Figs. \ref{frequency} (a)-(f) show the angular power spectrum and delay spread spectrum of three material surfaces at the three frequencies. It can be first observed that the type of material has a significant impact on both the power angular spectrum and the delay spread angular spectrum. Regarding the influence of frequency, as shown in Fig. \ref{frequency} (a)-(c), the received power of these three materials at the three frequencies is all concentrated in the specular reflection direction, exhibiting a certain degree of angular broadening. 
Moreover, compared with the power angular spectrum at 28 GHz, those at 8 GHz and 12 GHz show obvious similarity. Specifically, the received power at 8 GHz and 12 GHz is much stronger than that at 28 GHz.
Figs. \ref{frequency} (d)-(f) indicate that, at all the three frequencies, the angular spread of PDP is more pronounced in the incident plane. In addition, the delay spread at 28 GHz is quite different from (much larger than) that at the other two frequencies. However, for the two frequencies in the FR3 band, the change from 8 GHz to 12 GHz has minimal impact on the scattering power and delay spread. So in the subsequent discussion on the fitting performance of the scattering model for the FR3 band, we only present the fitting results from the 8 GHz scattering measurement.

% \begin{figure}[!t]
%     \centering
%     \begin{subfigure}{0.22\textwidth}
%         \centering
%         \includegraphics[width=\linewidth]{marblepower.png}
%         \caption{Marble wall power.}

%     \end{subfigure}%
%     \hfill
%     \begin{subfigure}{0.22\textwidth}
%         \centering
%         \includegraphics[width=\linewidth]{wenbopower.png}
%         \caption{Smooth wall power.}
   
%     \end{subfigure}
%     \hfill
%     \begin{subfigure}{0.22\textwidth}
%         \centering
%         \includegraphics[width=\linewidth]{brickpower.png}
%         \caption{Brick wall power.}

%     \end{subfigure}%
%     \hfill
%     \begin{subfigure}{0.22\textwidth}
%         \centering
%         \includegraphics[width=\linewidth]{marbledelay.png}
%         \caption{Marble wall delay spread.}

%     \end{subfigure}%
%     \hfill
%     \begin{subfigure}{0.22\textwidth}
%         \centering
%         \includegraphics[width=\linewidth]{wenbodelay.png}
%         \caption{Smooth wall delay spread.}
  
%     \end{subfigure}%
%     \hfill
%     \begin{subfigure}{0.22\textwidth}
%         \centering
%         \includegraphics[width=\linewidth]{brickdelay.png}
%         \caption{Brick wall delay spread.}
 
%     \end{subfigure}
%     \caption{(a)-(c) Angular spectrum of received power for three different surfaces at both 8GHz and 12GHz, (d)-(f) Angular spectrum of the delay spread of the PDP for these surfaces at the same two frequency bands.}
%     \label{frequency}
% \end{figure}

\begin{figure}[!t]  % 单列模板无需figure*，用figure即可
    \centering
    % 第1行
    \begin{subfigure}{0.48\textwidth}  % 子图宽度调整为0.48，两列更紧凑
        \centering
        \includegraphics[width=0.95\linewidth]{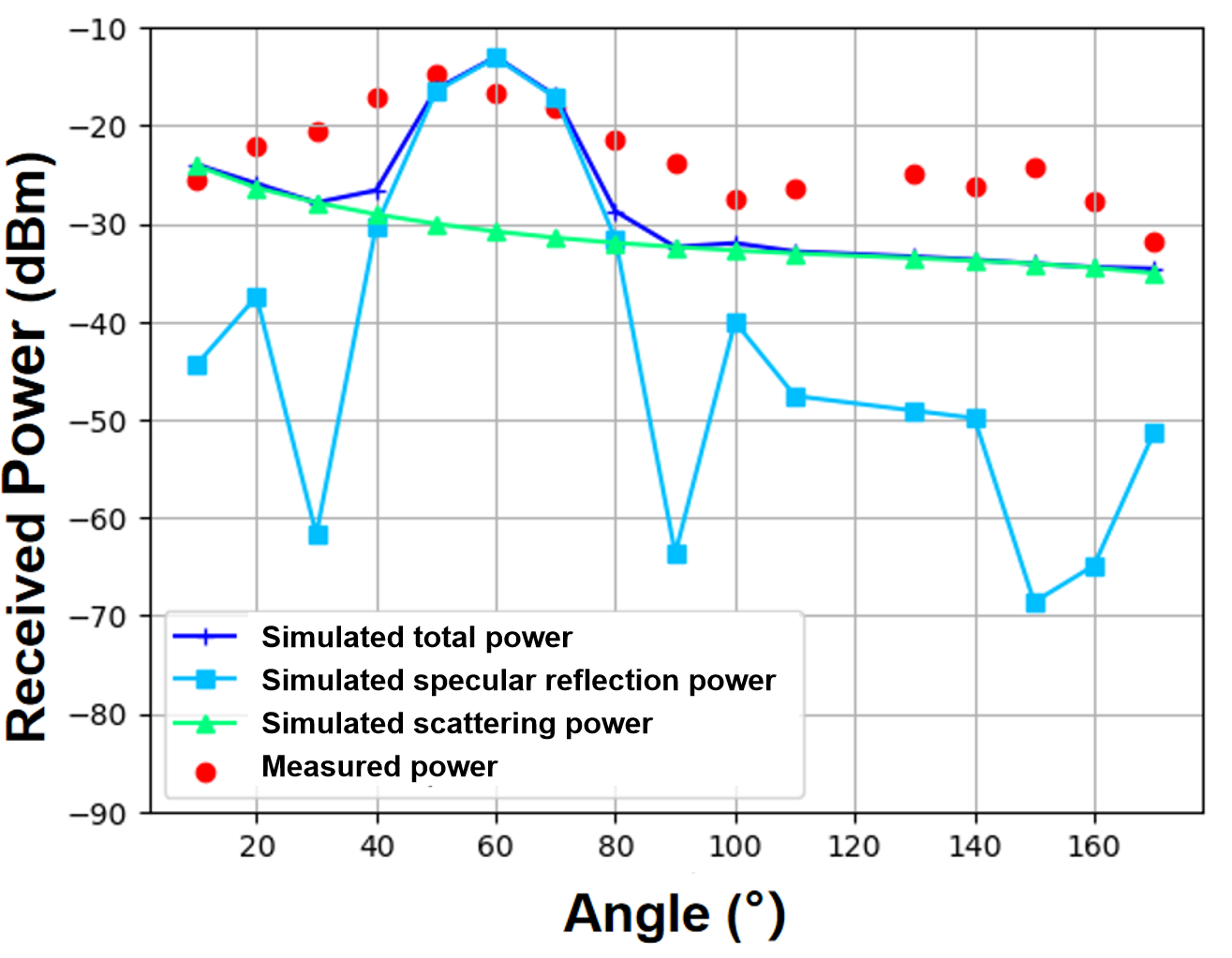}  % 图片占满子图空间
        \caption{Marble wall power (ER model).}  % 简化caption减少高度
    \end{subfigure}
    \hfill  % 两列间留白
    \begin{subfigure}{0.48\textwidth}
        \centering
        \includegraphics[width=0.95\linewidth]{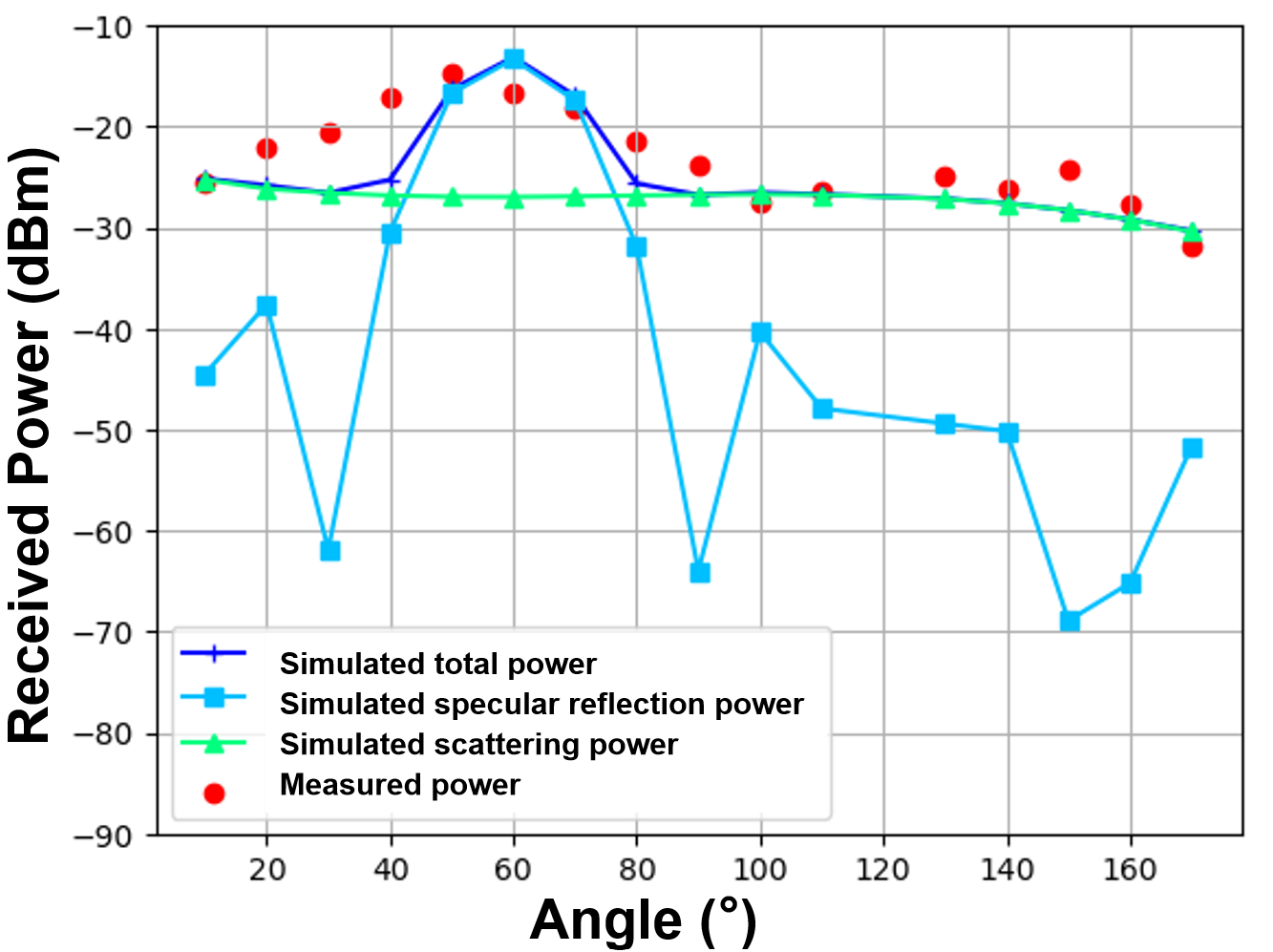}
        \caption{Marble wall power (BK model).}
    \end{subfigure}
    \\[0.5ex]  % 行间距（可微调，单位ex更适配字体）

    % % 第2行
    % \begin{subfigure}{0.48\textwidth}
    %     \centering
    %     \includegraphics[width=0.95\linewidth]{marbledelayER.png}
    %     \caption{Marble wall delay (ER model).}
    % \end{subfigure}
    % \hfill
    % \begin{subfigure}{0.48\textwidth}
    %     \centering
    %     \includegraphics[width=0.95\linewidth]{marbledelayBK.png}
    %     \caption{Marble wall delay (BK model).}
    % \end{subfigure}
    % \\[0.5ex]

    % 第3行
    \begin{subfigure}{0.48\textwidth}
        \centering
        \includegraphics[width=0.95\linewidth]{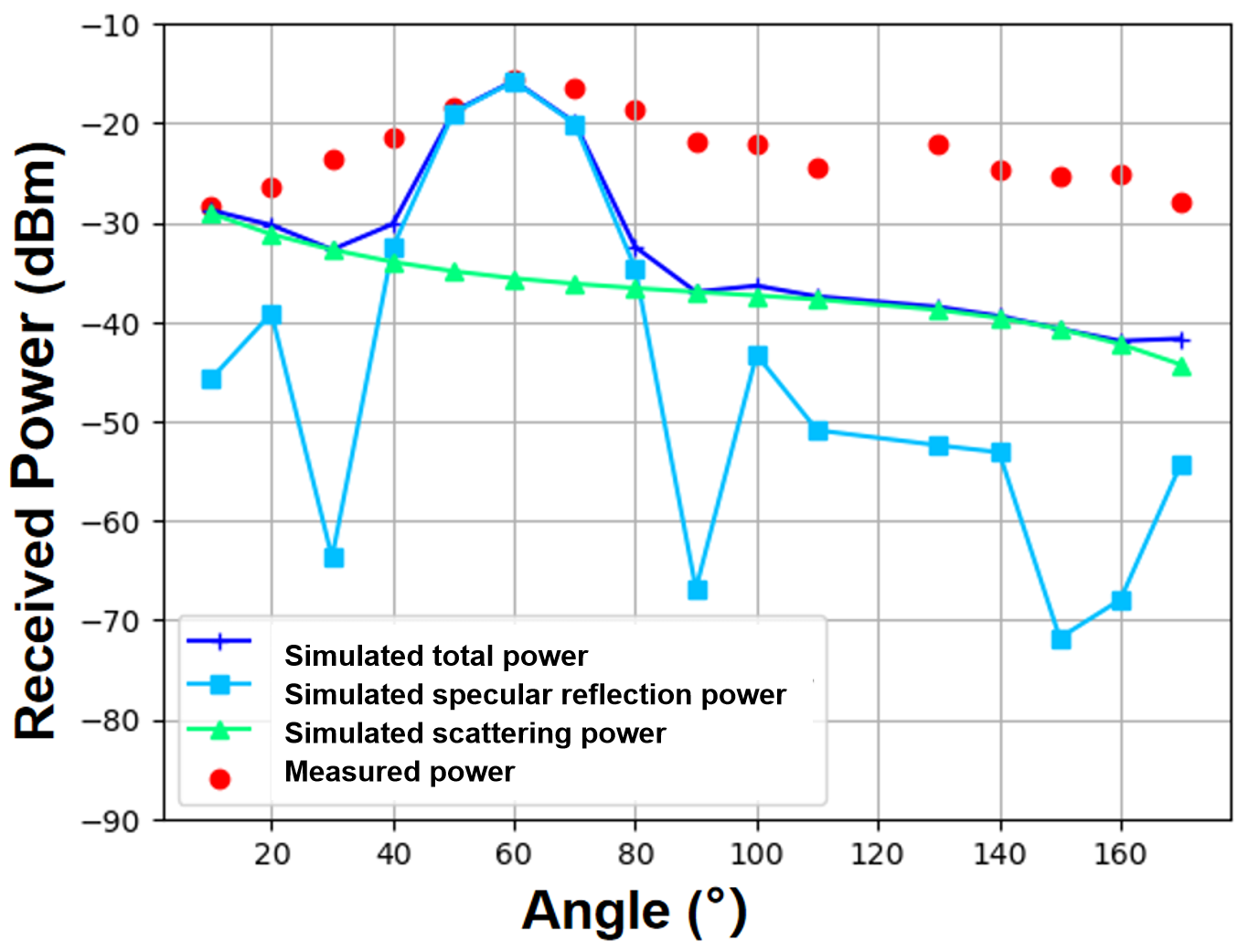}
        \caption{Smooth wall power (ER model).}
    \end{subfigure}
    \hfill
    \begin{subfigure}{0.48\textwidth}
        \centering
        \includegraphics[width=0.95\linewidth]{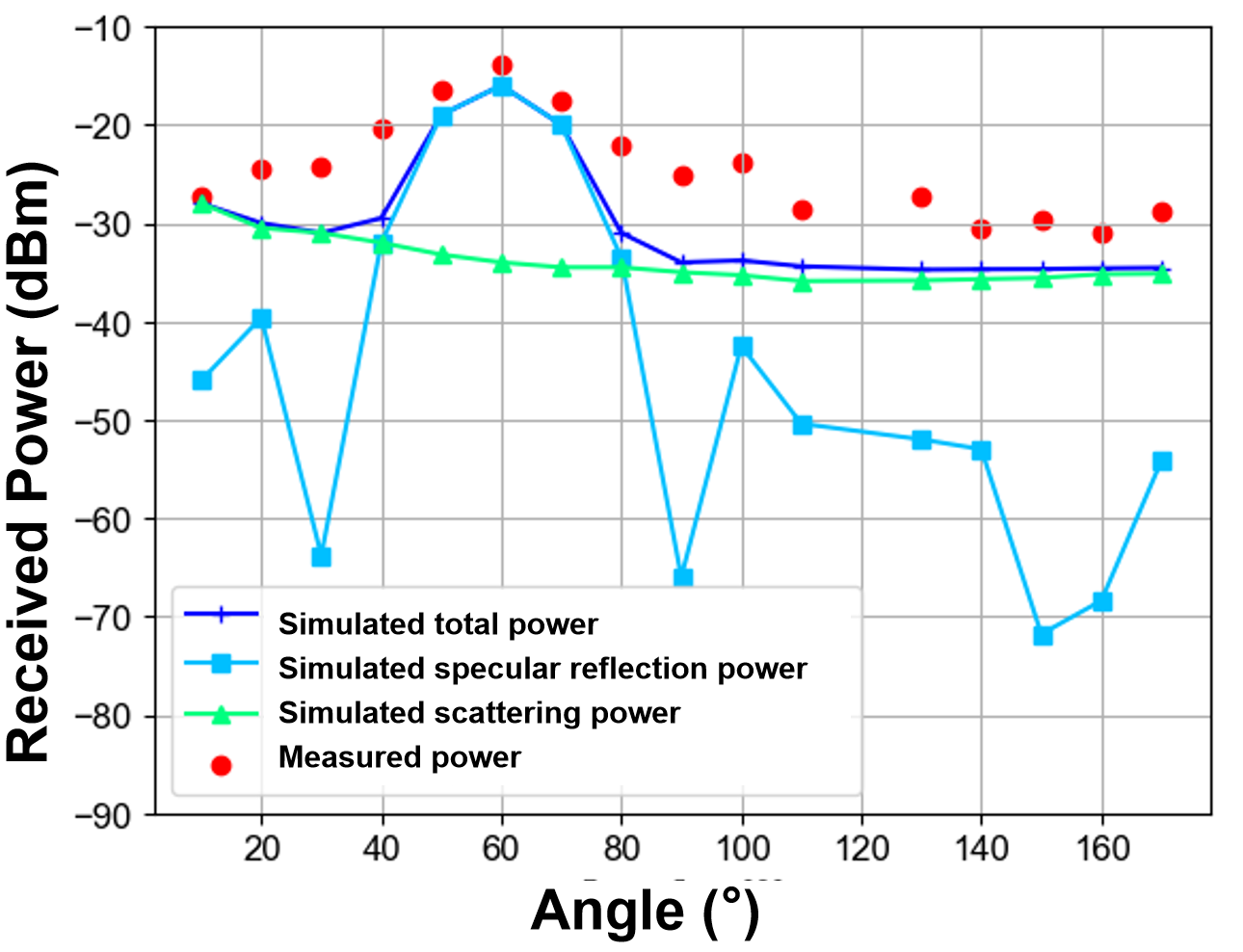}
        \caption{Smooth wall power (BK model).}
    \end{subfigure}
    \\[0.5ex]

    % % 第4行
    % \begin{subfigure}{0.48\textwidth}
    %     \centering
    %     \includegraphics[width=0.95\linewidth]{wenbodelayER.png}
    %     \caption{Smooth wall delay (ER model).}
    % \end{subfigure}
    % \hfill
    % \begin{subfigure}{0.48\textwidth}
    %     \centering
    %     \includegraphics[width=0.95\linewidth]{wenbodelayBK.png}
    %     \caption{Smooth wall delay (BK model).}
    % \end{subfigure}
    % \\[0.5ex]

    % 第5行
    \begin{subfigure}{0.48\textwidth}
        \centering
        \includegraphics[width=0.95\linewidth]{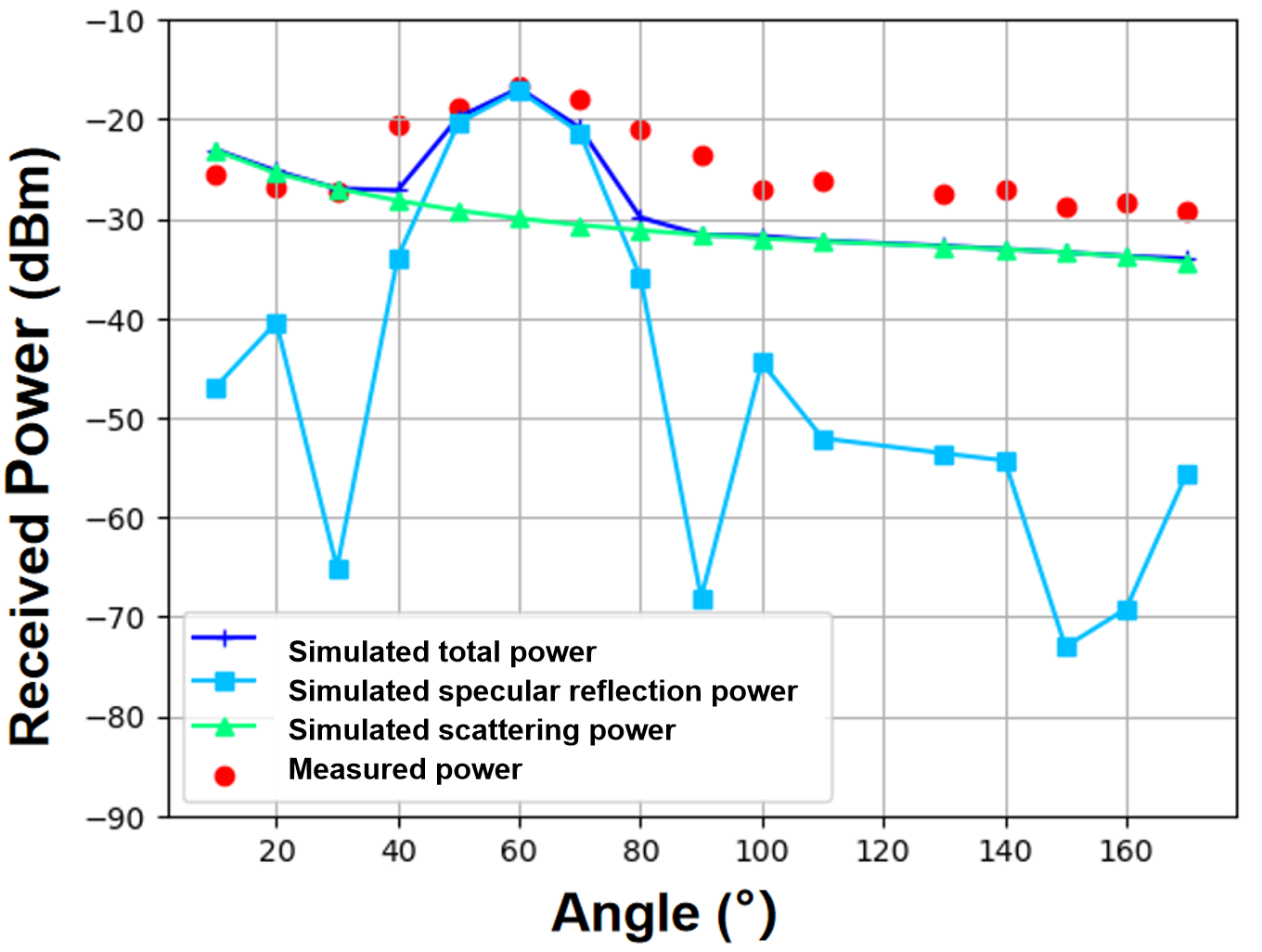}
        \caption{Brick wall power (ER model).}
    \end{subfigure}
    \hfill
    \begin{subfigure}{0.48\textwidth}
        \centering
        \includegraphics[width=0.95\linewidth]{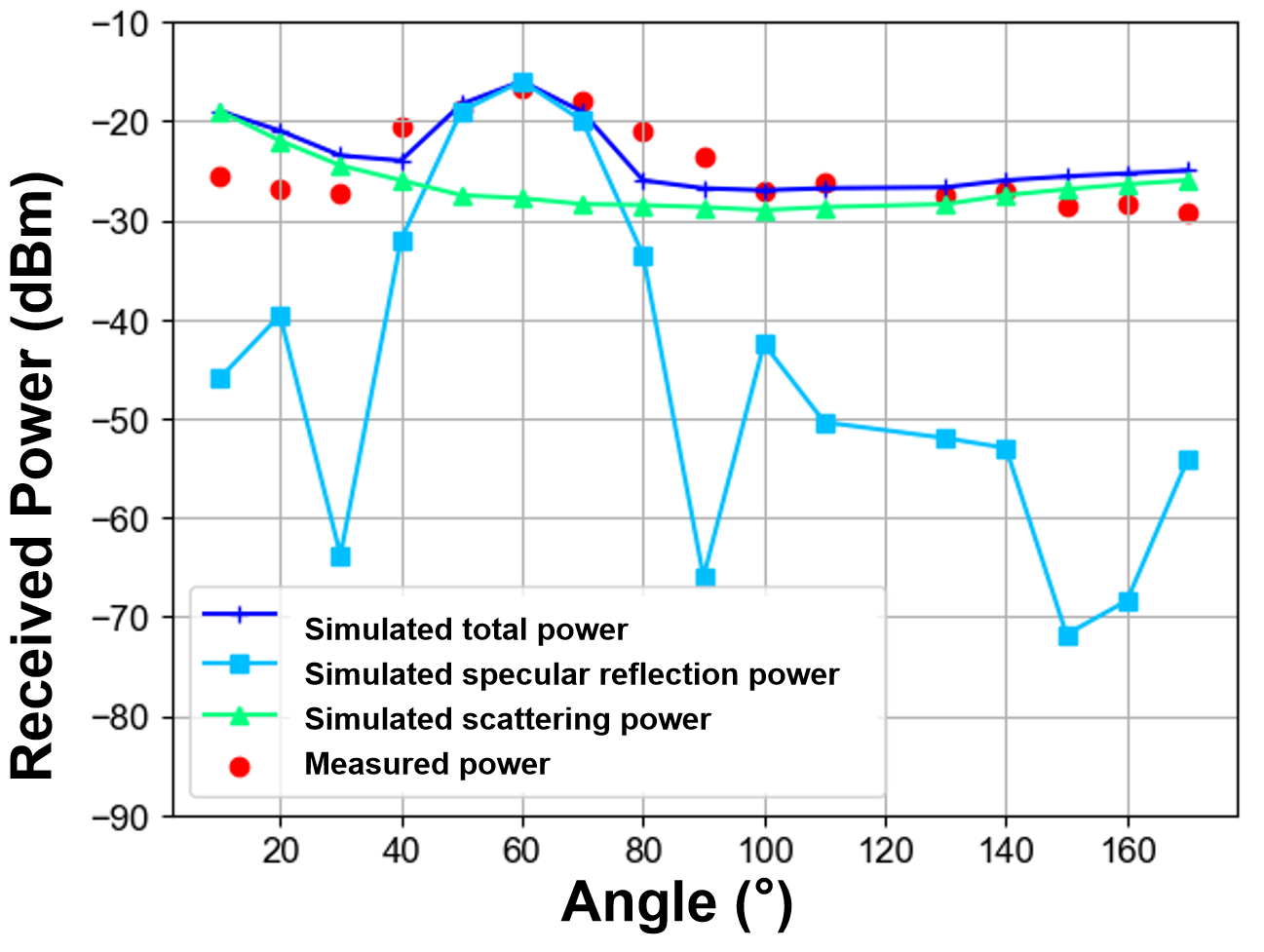}
        \caption{Brick wall power (BK model).}
    \end{subfigure}
    \\[0.5ex]

    % % 第6行
    % \begin{subfigure}{0.48\textwidth}
    %     \centering
    %     \includegraphics[width=0.95\linewidth]{brickdelayER.png}
    %     \caption{Brick wall delay (ER model).}
    % \end{subfigure}
    % \hfill
    % \begin{subfigure}{0.48\textwidth}
    %     \centering
    %     \includegraphics[width=0.95\linewidth]{brickdelayBK.png}
    %     \caption{Brick wall delay (BK model).}
    % \end{subfigure}

    \caption{(a), (b): Marble wall; (c), (d) Smooth wall; (e)-(f): Brick wall. (ER/BK model fitting results for power data.)}  % 简化总caption
    \label{powercom}
\end{figure}

\begin{figure}[!t]  % 单列模板无需figure*，用figure即可
    \centering
    % % 第1行
    % \begin{subfigure}{0.48\textwidth}  % 子图宽度调整为0.48，两列更紧凑
    %     \centering
    %     \includegraphics[width=0.95\linewidth]{marblepowerER.png}  % 图片占满子图空间
    %     \caption{Marble wall power (ER model).}  % 简化caption减少高度
    % \end{subfigure}
    % \hfill  % 两列间留白
    % \begin{subfigure}{0.48\textwidth}
    %     \centering
    %     \includegraphics[width=0.95\linewidth]{marblepowerBK.png}
    %     \caption{Marble wall power (BK model).}
    % \end{subfigure}
    % \\[0.5ex]  % 行间距（可微调，单位ex更适配字体）

    % 第2行
    \begin{subfigure}{0.48\textwidth}
        \centering
        \includegraphics[width=0.95\linewidth]{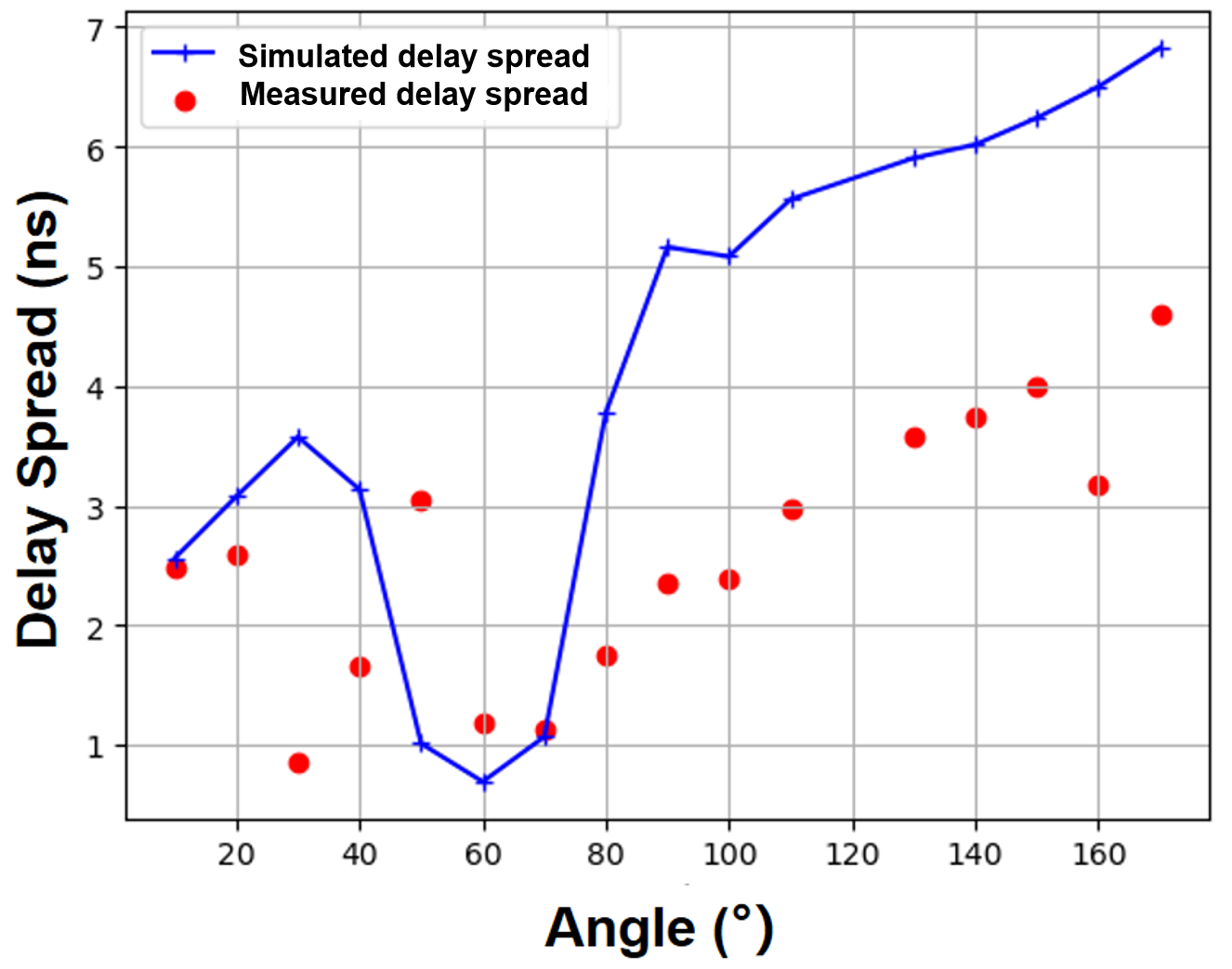}
        \caption{Marble wall delay (ER model).}
    \end{subfigure}
    \hfill
    \begin{subfigure}{0.48\textwidth}
        \centering
        \includegraphics[width=0.95\linewidth]{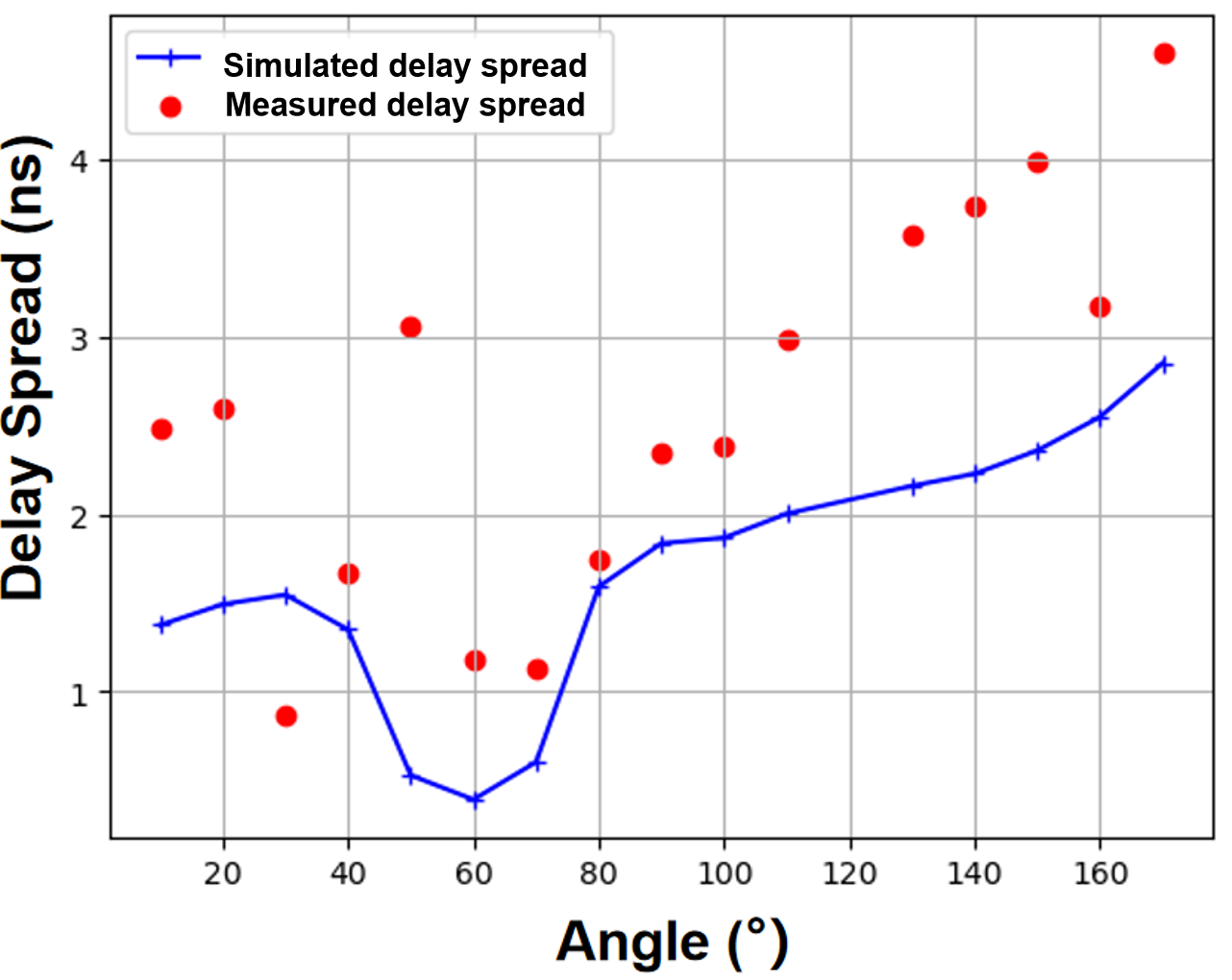}
        \caption{Marble wall delay (BK model).}
    \end{subfigure}
    \\[0.5ex]

    % % 第3行
    % \begin{subfigure}{0.48\textwidth}
    %     \centering
    %     \includegraphics[width=0.95\linewidth]{wenbopowerER.png}
    %     \caption{Smooth wall power (ER model).}
    % \end{subfigure}
    % \hfill
    % \begin{subfigure}{0.48\textwidth}
    %     \centering
    %     \includegraphics[width=0.95\linewidth]{wenbopowerBK.png}
    %     \caption{Smooth wall power (BK model).}
    % \end{subfigure}
    % \\[0.5ex]

    % 第4行
    \begin{subfigure}{0.48\textwidth}
        \centering
        \includegraphics[width=0.95\linewidth]{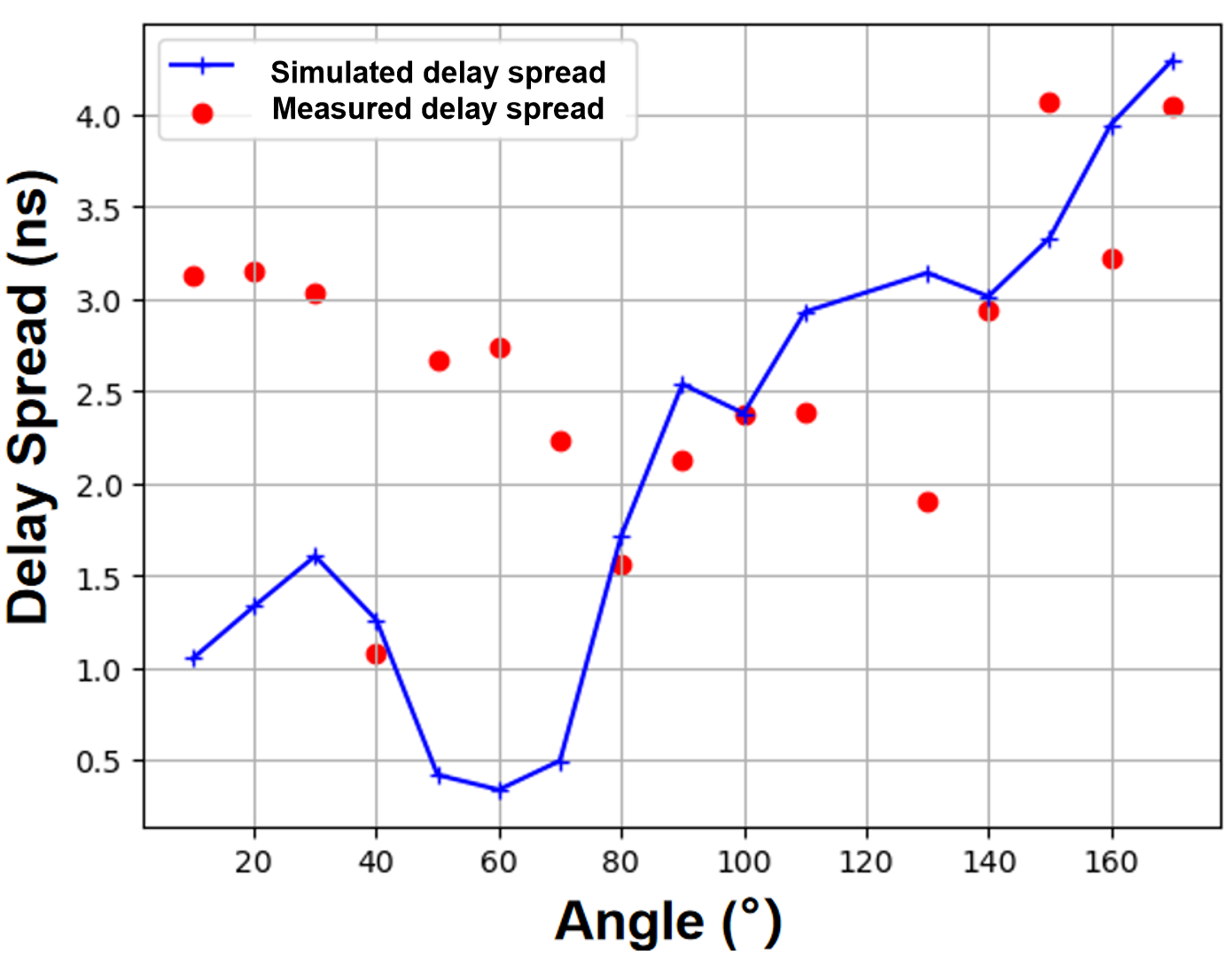}
        \caption{Smooth wall delay (ER model).}
    \end{subfigure}
    \hfill
    \begin{subfigure}{0.48\textwidth}
        \centering
        \includegraphics[width=0.95\linewidth]{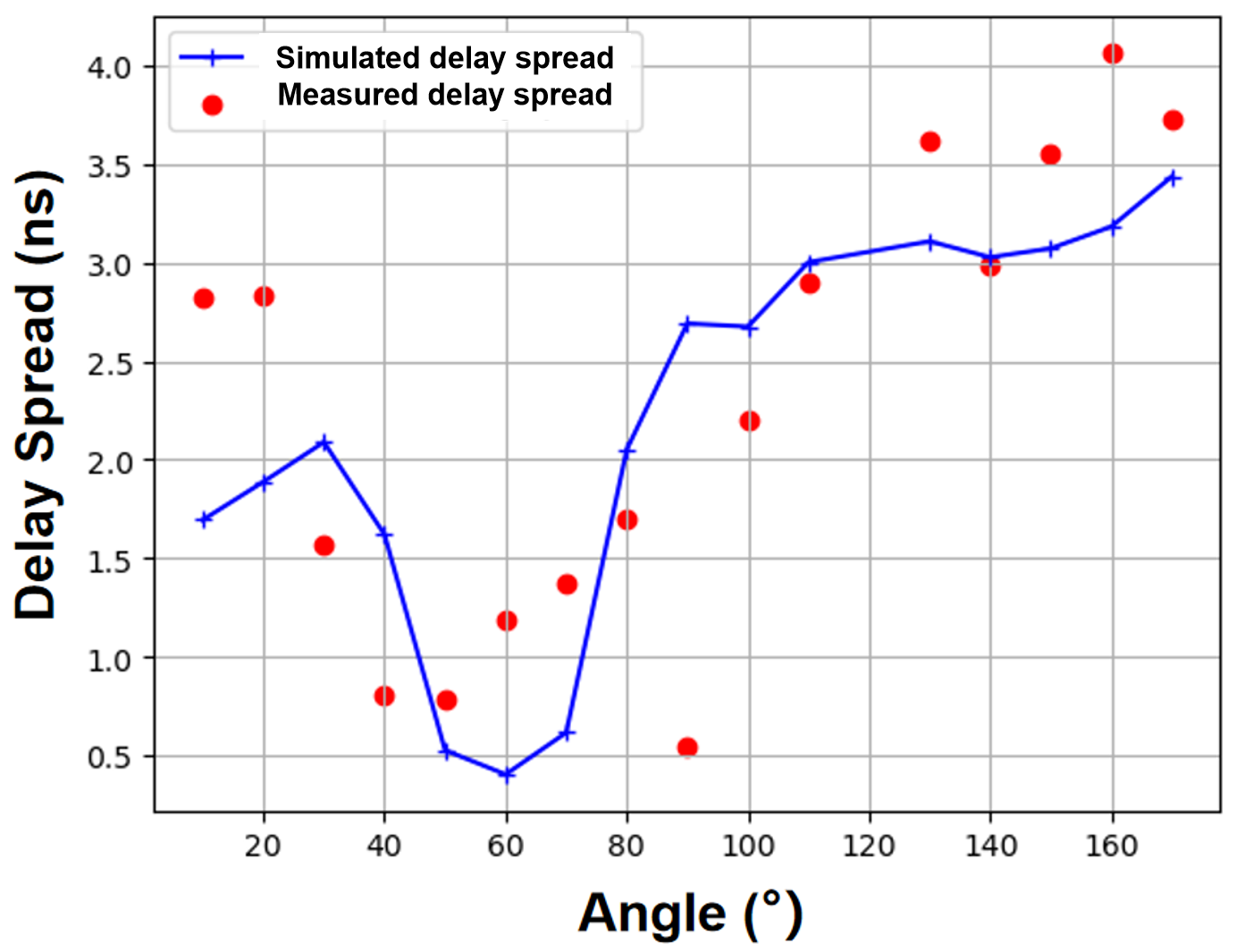}
        \caption{Smooth wall delay (BK model).}
    \end{subfigure}
    \\[0.5ex]

    % % 第5行
    % \begin{subfigure}{0.48\textwidth}
    %     \centering
    %     \includegraphics[width=0.95\linewidth]{brickpowerER.png}
    %     \caption{Brick wall power (ER model).}
    % \end{subfigure}
    % \hfill
    % \begin{subfigure}{0.48\textwidth}
    %     \centering
    %     \includegraphics[width=0.95\linewidth]{brickpowerBK.png}
    %     \caption{Brick wall power (BK model).}
    % \end{subfigure}
    % \\[0.5ex]

    % 第6行
    \begin{subfigure}{0.48\textwidth}
        \centering
        \includegraphics[width=0.95\linewidth]{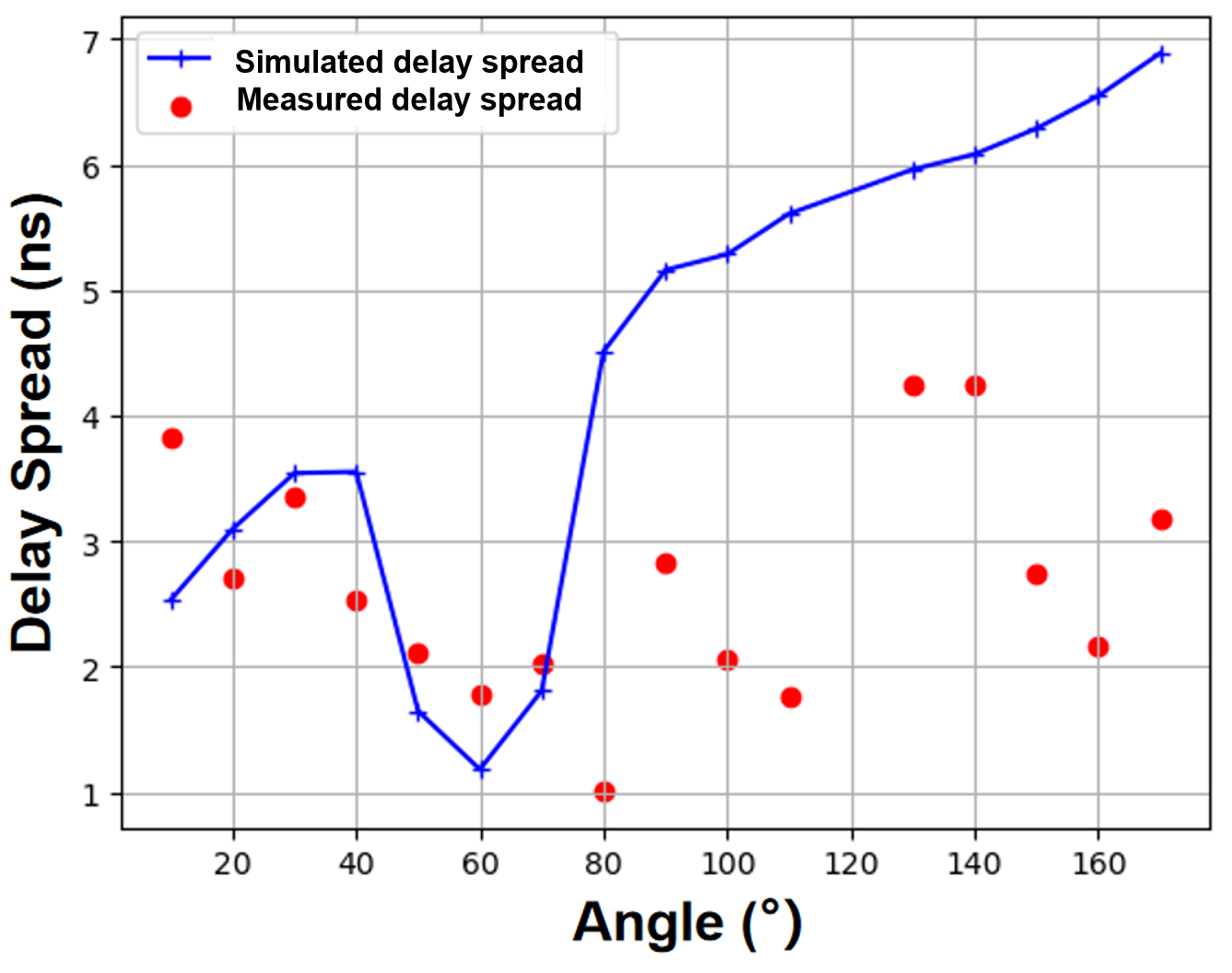}
        \caption{Brick wall delay (ER model).}
    \end{subfigure}
    \hfill
    \begin{subfigure}{0.48\textwidth}
        \centering
        \includegraphics[width=0.95\linewidth]{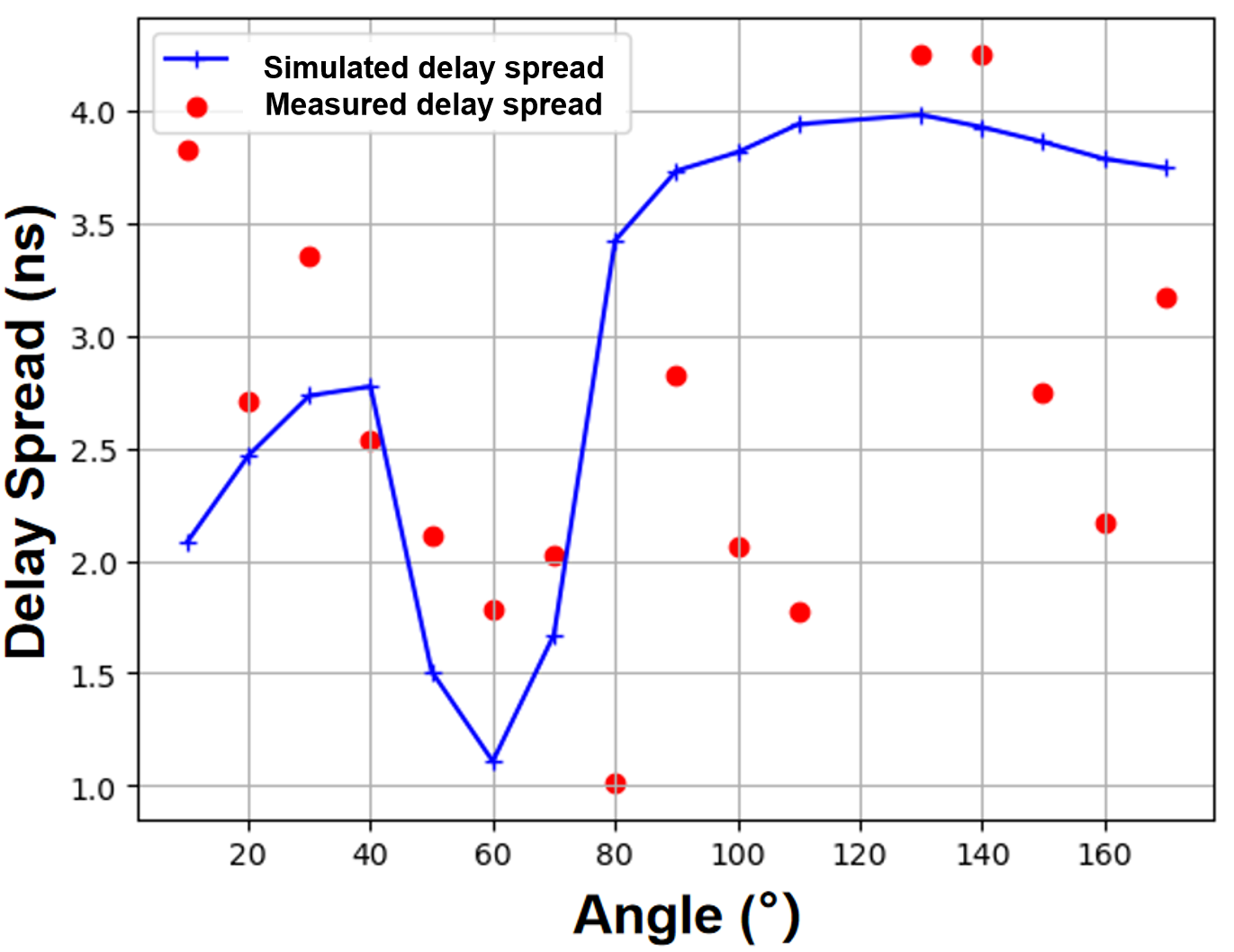}
        \caption{Brick wall delay (BK model).}
    \end{subfigure}

 \caption{(a), (b): Marble wall; (c), (d) Smooth wall; (e)-(f): Brick wall. (ER/BK model fitting results for delay spread data.)}  % 简化总caption
    \label{delayspreadcom}
\end{figure}

\subsection{Fitting Results Using ER and BK Models }

We fitted the received power angular spectrum and delay spread angular spectrum data for the three material at 8 GHz using both the ER model and the BK model (using planar data for marble walls and smooth walls, and data from four receiver heights for brick walls).

For the brick wall, the backscattering lobe model within the ER model was employed. Fig. \ref{powercom} and Fig. \ref{delayspreadcom} respectively present the best fitting results of two parameters, received power and delay spread, under the ER and BK models for three materials, while Table \ref{model_fitting} presents the parameters of the two scattering models for the three surfaces. It is evident that when simultaneously considering the energy angular spectrum and the delay spread angular spectrum, the ER model does not fit the measurement data well, whereas the BK model demonstrates better adaptability in fitting both the energy spectrum and the delay spectrum.
\begin{figure}[!t]
    \centering
    \includegraphics[width=0.8\columnwidth]
    {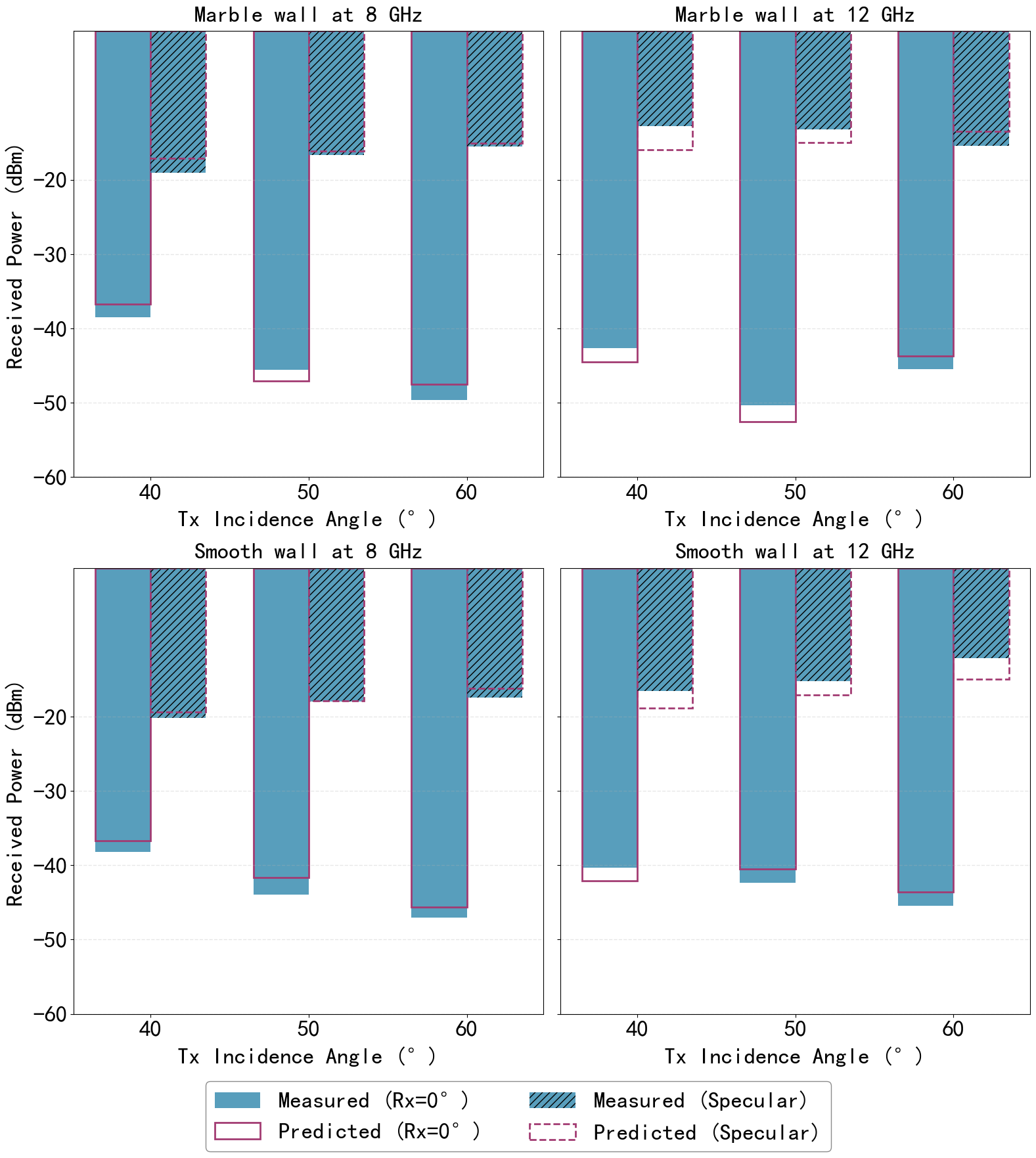}
    \caption{\textcolor{black}{Comparison between BK model predictions and measured values of power at non-30° incidence angles.}}
    \label{val}
\end{figure}

\subsection{\textcolor{black}{Evaluation of the generalization capability of the parameterized model under different incidence angles}
}

\textcolor{black}{The parameterized BK model is capable of predicting the angle spectrum of received power under arbitrary incidence angles. Using the aforementioned parameterized BK model, scattering simulations were performed for two types of surfaces (marble and smooth wall) at two frequencies (8 GHz and 12 GHz) with incidence angles of 40°, 50°, and 60°. As illustrated in Fig. \ref{val}, a comparison between the simulated results and the measured power values shows that the model parameters fitted under the 30 ° incidence angle can effectively predict the scattering power under other incidence angles. This shows that the model, parameterized using measured data from a single incidence angle, can be generalized to scenarios involving other incidence angles.}

\subsection{Parameterization results of the ER-BK hybrid model}
% \begin{figure}[!t]
%     \centering
%     \includegraphics[width=0.8\columnwidth]
%     {fusio.png}
%     \caption{ER model fitting result for BK simulation data, 28 GHz, marble ($h_{\text{rms}} = 1.1 \text{mm}, T = 5\text{mm}$).}
%     \label{fusion}
% \end{figure}
\begin{figure}[!t]
    \centering
    \includegraphics[width=0.8\columnwidth]
    {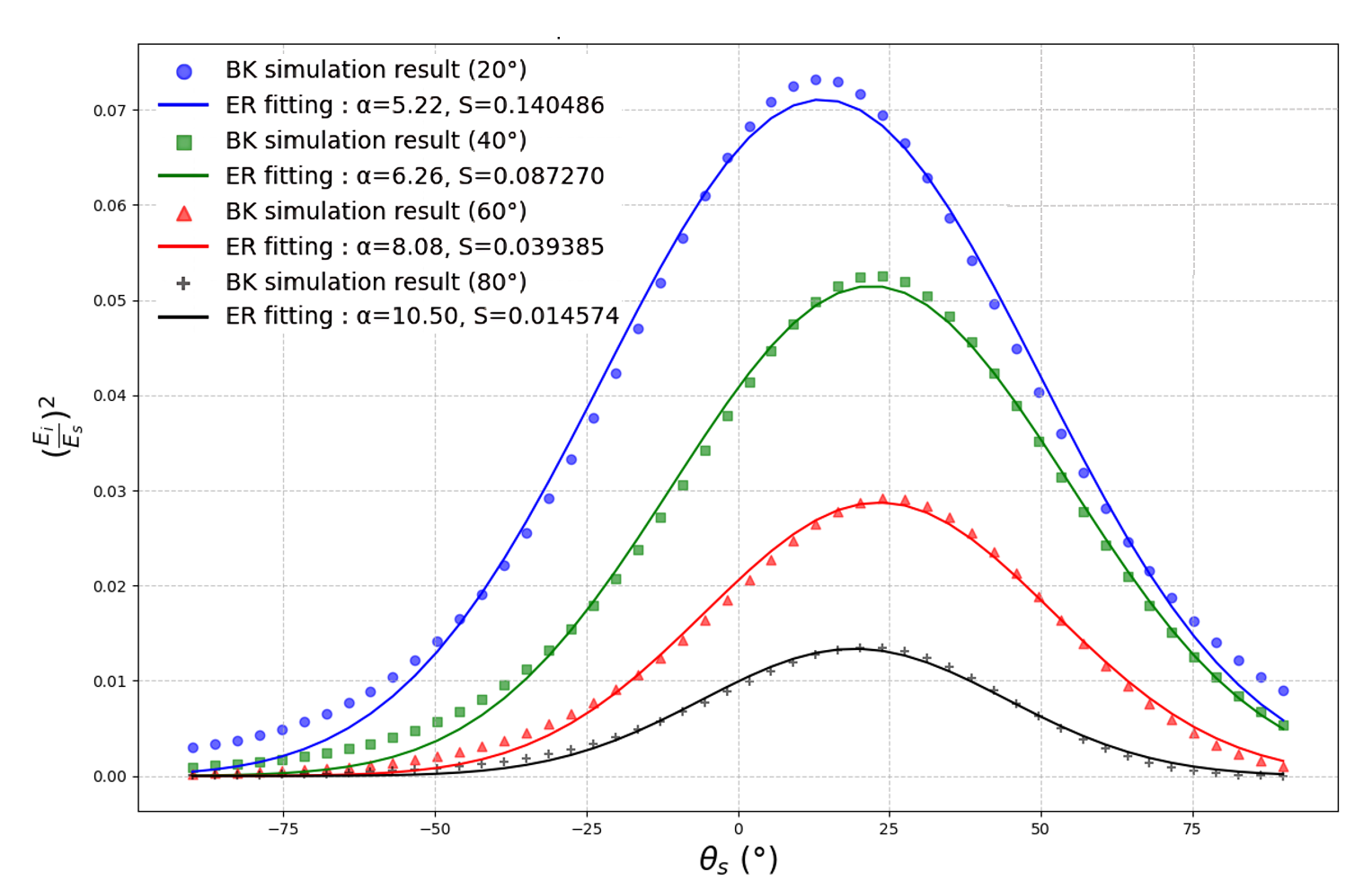}
    \caption{ER model fitting result for BK simulation data, 28 GHz, marble ($h_{\text{rms}} = 1.1 \text{mm}, T = 5\text{mm}$).}
    \label{fusion}
\end{figure}
Following the ER-BK hybrid model introduced in Section before, we conducted BK model simulations at 28 GHz using the BK model fitting parameters derived from the above 8 GHz measurement data. Fig \ref{fusion} shows the BK scattering patterns of marble material and the corresponding parameter fitting results of the improved ER model under three incident angles (20°, 40°, 60°, and 80°) at 28 GHz. 

First, it can be observed that in our proposed method, the improved ER model can effectively fit the simulated scattering patterns of the BK model. Second, we can observe the variation patterns of some parameters of the ER model with the incident angle. It can be seen that the larger the incident angle, the smaller the scattering coefficient $S$, which is consistent with the conventional ER model. Meanwhile, beamwidth $\alpha$ increases as the incident angle becomes larger, meaning that the lobe widens with the increase of the incident angle. This is a phenomenon that cannot be described by the traditional ER model. \textcolor{black}{One plausible physical mechanism behind it is that as the incident angle increases, the illuminated area expands; this causes the phase of electromagnetic waves to become more disordered after interacting with the surface, ultimately leading to a more dispersed spatial distribution of the waves.}

\section{Discussion}

% In this study, we explore the fundamental diffuse scattering mechanisms at representative centimeter-wave and millimeter-wave frequency bands. Diffuse scattering measurements and model parameterization were conducted on three typical building wall surfaces at 28 GHz millimeter-wave frequency, as well as at two frequencies (8 GHz and 12 GHz) within the FR3 band. In terms of measurement, a 3D measurement scheme is proposed to provide spatial information for diffuse scattering data. Regarding the parameterization method, we put forward a parameterization approach for the scattering model by utilizing two dimensions (total power and delay spread) of high-bandwidth PDP data. In terms of the model, the ER-BK hybrid model is proposed, which integrates the accuracy of the BK model and the simplicity of the ER model.

% Using the method we propose,
% first, the 3D measurement procedure was validated to be effective in enhancing parameterization accuracy through the data at 28 GHz. Second, the analysis of the impact of frequency and surface materials on diffuse scattering measurement results reveals that the diffuse scattering data at 8 GHz and 12 GHz are highly similar in comparison with that at 28 GHz. Third, a comparison of the fitting effects of the ER model and the BK model on diffuse scattering at 8 GHz and 12 GHz within FR3 demonstrated that the BK model is more advantageous in the measurement data. Finally, the simulation of the parameterized marble surface using the ER-BK hybrid model confirmed the feasibility of this method.

\textcolor{black}{In this study, we investigate diffuse scattering mechanisms across centimeter-wave and millimeter-wave bands, performing measurements and model parameterization on three typical building wall surfaces. measurements cover the 28 GHz millimeter-wave frequency and two frequencies within the FR3 band (8 GHz and 12 GHz). Specifically, we propose three key improvements for building surface scattering measurement and modeling:
Firstly, we introduce a 3D measurement scheme to capture spatial scattering data. This enables the extraction of comprehensive spatial scattering information, a significant enhancement over the limitations of traditional 2D angular measurements \cite{ju2019scattering}, \cite{pascual2016importance}, \cite{guo2024diffuse}.
Secondly, we utilize a parameterization approach based on total power and delay spread from high-bandwidth PDP. In contrast to conventional methods that rely only on fitting the angular power spectrum (e.g., \cite{4052607}, \cite{pascual2016importance}, \cite{tian2019effect}), our technique exploits the full temporal richness of high-bandwidth PDPs to derive more accurate model parameters.
Finally, we introduce an ER-BK hybrid model that integrates the accuracy of the BK model and the simplicity of the ER model, thereby offering a novel and balanced approach to surface scattering modeling.}

\textcolor{black}{Our results yield several insights: First, the 3D measurement scheme is validated to enhance parameterization accuracy, confirming its effectiveness in capturing spatial scattering features. Second, diffuse scattering characteristics at 8 GHz and 12 GHz are found to be highly similar,blue{which carries important implications for FR3 band channel modeling—it may support the development of a unified or simplified model framework for this frequency range, reducing modeling complexity.} Third, the BK model demonstrates superior fitting performance for FR3 band data, and notably, models parameterized under a single incident angle exhibit good generalization ability for predicting scattering at other angles, simplifying the parameterization process. Finally, the ER-BK hybrid model proves feasible for simulating parameterized surfaces, balancing accuracy and computational efficiency. }

\textcolor{black}{However, this work still has two limitations: First, environmental factors (e.g., temperature, humidity) were not considered in scattering parameterization, which are critical for real-world applicability. Second, the computational complexity of the existing model remains non-trivial, posing challenges for efficient deployment in large-scale scenarios. Corresponding future research directions address these limitations: First, conduct measurements on surfaces under different environments (e.g., humidity), investigate the impact of environmental factors, and incorporate them into the model. Second, introduce AI-driven methods to improve modeling efficiency and accuracy, or develop hybrid physics-data models that combine physical interpretability  with data-driven fitting capability, further boosting modeling performance.
}

% \section{\textcolor{black}{Data Availability}}
% \textcolor{blue}{We do not have any research data declarations to make.}
% \section{\textcolor{blue}{Competing Interests}}
% \textcolor{blue}{We have no competing interests as defined by Nature Portfolio, or other interests that might be perceived to influence the results and/or discussion reported in this paper.} 
% \section{\textcolor{black}{Author Contribution}}
% \textcolor{blue}{T. Z. conducted measurements, wrote the main manuscript text, and prepared all the figures. S. S. designed the study, developed research methods, wrote the main manuscript text, and secured financial support. M. T. supervised the project and secured financial support.  T. Z.,  S. S.,  M. T., Q. Z. and R. G. reviewed and revised the manuscript.}
% \section{\textcolor{black}{Acknowledgement}}
% \textcolor{blue}{This work is supported by the National Natural Science Foundation of China under Grants 62271310, 62431014, 62125108, 62271250, and 62571276. We would also like to express our sincere gratitude to Mr. Yulu Guo for his invaluable help and technical support during the diffuse scattering measurement campaign. His support in the measurement setup and data collection was essential to this work.}

% \bibliographystyle{plain}
\bibliography{references}

\vspace{12pt}

\end{document}